%% file: main.tex
\documentclass[reprint,
preprintnumbers,
nofootinbib,
amsmath,amssymb,
aps,
prd,
]{revtex4-2}

\usepackage{graphicx}
\usepackage{dcolumn}
\usepackage{bm}
\usepackage[hidelinks]{hyperref}
\usepackage[all]{hypcap}
\usepackage{xcolor}
\usepackage{enumitem}
\usepackage{animate}
\usepackage{multirow}

\hypersetup{
	colorlinks,
	linkcolor={red!50!black},
	citecolor={blue!50!black},
	urlcolor={blue!80!black}
}

\DeclareMathOperator{\Tr}{Tr}

\makeatletter
\def\maketitle{
	\@author@finish
	\title@column\titleblock@produce
	\suppressfloats[t]}
\makeatother

\begin{document}

\preprint{ADP-24-09/T1248}

\title{Centre vortex geometry at finite temperature}

\author{Jackson A. Mickley}
\author{Waseem Kamleh}
\author{Derek B. Leinweber}
\affiliation{Centre for the Subatomic Structure of Matter, Department of Physics, The University of Adelaide, South Australia 5005, Australia}

\begin{abstract}
The geometry of centre vortices is studied in SU(3) gauge theory at finite temperature to capture the key structural changes that occur through the deconfinement phase transition. Visualisations of the vortex structure in temporal and spatial slices of the lattice reveal a preference for the vortex sheet to align with the temporal dimension above the critical temperature. This is quantified through a correlation measure. A collection of vortex statistics, including vortex and branching point densities, and vortex path lengths between branching points, are analysed to highlight internal shifts in vortex behaviour arising from the loss of confinement. We find the zero-temperature inclination of branching points to cluster at short distances vanishes at high temperatures, embodying a rearrangement of branching points within the vortex structure. These findings establish the many aspects of centre vortex geometry that characterise the phase transition in pure gauge theory.
\end{abstract}

\maketitle

\section{Introduction} \label{sec:intro}
The centre vortex picture \cite{VortexI,VortexII,VortexIII,ConfinementReview} is well established as a prime candidate for underpinning many of the emergent phenomena in quantum chromodynamics (QCD), including dynamical chiral symmetry breaking \cite{CSBI,CSBII,CSBIII,CSBIV,CSBV,CSBVI,CSBVII} and confinement \cite{CSBI,CSBV,ConfinementI,ConfinementII,ConfinementIII,ConfinementIV,CasimirScalingI,CasimirScalingII,ConfinementV,PvortexStructure,ConfinementVI,ConfinementVII,ConfinementVIII,ConfinementIX,ConfinementX,ConfinementXI,ConfinementXII,ConfinementDynamicalI,ConfinementDynamicalII}. Centre vortices naturally give rise to an area-law falloff for large Wilson loops \cite{VortexAreaLawI,VortexAreaLawII},
\begin{equation} \label{eq:arealaw}
	\langle W(C) \rangle \sim \exp\left(-\sigma A(C)\right) \,,
\end{equation}
with such behaviour signifying confinement for static heavy quarks \cite{ConfinementAreaLawI,ConfinementAreaLawII,ConfinementAreaLawIII,ConfinementReview}. Due to string breaking at large separations \cite{StringBreaking}, in the presence of light quarks one can turn to the gluon propagator, where if there is no K{\"a}llen-Lehmann representation, the corresponding physical states are confined. This is often inferred by studying the Schwinger function of the gluon propagator, for which negative values at large Euclidean times signal positivity violation in the spectral density function \cite{PositivityViolationI,PositivityViolationII}.

In SU(2) Yang-Mills theory, centre vortices have been demonstrated to account for 100\% of the string tension $\sigma$ \cite{ConfinementI,ConfinementIV}, though full recovery in pure SU(3) theory has proven more challenging. Despite vortex removal resulting in a vanishing string tension, static quark potential calculations on vortex-only fields has shown to recover only $\approx 62\%$ of the string tension in SU(3) \cite{ConfinementXI}. Remarkably, recent work considering the impact of dynamical fermions on centre vortices for the first time showed that the string tension can be entirely recreated from centre vortices in full QCD \cite{ConfinementDynamicalI}. In addition, the vortex-removed fields display no signs of positivity violation with dynamical fermions, another advantage compared to the pure gauge theory which still exhibited a remnant effect at long distances \cite{ConfinementDynamicalII}. These findings reaffirm centre vortices as fundamental to the nonperturbative nature of confinement, though the inconsistencies in the pure gauge regime remain an open problem.

Following a phase transition at some critical temperature $T_c$, the SU($N$) vacuum is understood to exist in a deconfined state. This motivates exploring centre vortices at a range of temperatures through the deconfinement phase transition in SU(3) to identify the principal structural changes that can be attributed to confinement. We utilise previously developed visualisation techniques \cite{Visualisations} to qualitatively study the changes in vortex geometry. This is followed by a focused analysis into several vortex statistics as a function of temperature, such as vortex and branching point densities. Given they do not occur in SU(2), branching points provide a particularly interesting avenue of research, and vortex branching is known to experience substantial bulk changes at $T_c$ \cite{ConfinementXII,Branching}. We subsequently take this further by studying the distances between successive branching points and the accompanying branching probabilities, in turn searching for any internal rearrangement of branching points at high temperatures. In this regard, the pure gauge theory provides fertile ground for drawing out the salient features of vortex geometry at finite temperature that can guide future work with the inclusion of dynamical fermions in QCD.

This paper is structured as follows. In Sec.~\ref{sec:centrevortices}, the centre vortex model is briefly reviewed along with the procedure for identifying vortices on the lattice. Section \ref{sec:visualisations} contains visualisations of the vortex structure at a range of temperatures either side of $T_c$ and the accompanying discussion. Our detailed analysis on the evolution of intrinsic vortex statistics with temperature is presented throughout Secs.~\ref{sec:statistics} and \ref{sec:branchingpoints}. Finally, we summarise our main findings in Sec.~\ref{sec:conclusion}. Supplemental Material providing embedded animations of centre vortex structures is located at the end of this document. Instructions on interacting with these animations is given therein, and the figures are referenced in the main text as Fig.~S-x.

\section{Centre vortices} \label{sec:centrevortices}
Centre vortices, originally introduced in Ref.~\cite{VortexI}, are regions of the gauge field that carry magnetic flux quantised according to the centre of SU(3),
\begin{equation}
    \mathbb{Z}_3 = \left\{ \exp\left(\frac{2\pi i}{3}\, n \!\right) \mathbb{I} \;\middle|\; n = -1,0,1 \right\} \,.
\end{equation}
Physical vortices in the QCD ground-state fields have a finite thickness and so permeate all four spacetime dimensions. In contrast, on the lattice ``thin" centre vortices are extracted through a well-known gauge-fixing procedure that seeks to bring each link variable $U_\mu(x)$ as close as possible to an element of $\mathbb{Z}_3$, known as Maximal Centre Gauge (MCG). These thin vortices form closed surfaces in four-dimensional Euclidean spacetime, and thus one-dimensional structures in a three-dimensional slice of the four-dimensional spacetime.

Fixing to MCG is typically performed by finding the gauge transformation $\Omega(x)$ to maximise the functional \cite{MCG}
\begin{align}
    R = \frac{1}{V\,N_\mathrm{dim}\,N_c^2} \,\sum_{x,\,\mu} \,\left| \Tr U_\mu^{\Omega}(x) \right|^2 \,, && V = N_s^3 \times N_t \,.
\end{align}
The links are subsequently projected onto the centre,
\begin{equation}
	U_\mu(x) \longrightarrow Z_\mu(x) = \exp\left(\frac{2\pi i}{3} \,n_\mu(x) \!\right) \mathbb{I} \in \mathbb{Z}_3 \,,
\end{equation}
with $n_\mu(x) \in \{-1,0,1\}$ identified as the centre phase nearest to $\arg \Tr U_\mu(x)$ for each link. Finally, the locations of vortices are identified by nontrivial plaquettes in the centre-projected field,
\begin{equation}
	P_{\mu\nu}(x) = \prod_\square Z_\mu(x) = \exp\left(\frac{2\pi i}{3} \, m \!\right)\mathbb{I}
\end{equation}
with $m=\pm 1$. The value of $m$ is referred to as the \textit{centre charge} of the vortex, and we say the plaquette is pierced by a vortex.

Due to a Bianchi identity satisfied by the projected vortex fields \cite{ConfinementXII, Branching}, the centre charge is conserved such that the vortex topology manifests as closed sheets in four dimensions, or as closed lines in three-dimensional slices of the lattice. Although gauge-dependent, numerical evidence strongly suggests the projected vortices' locations are correlated with the physical ``thick'' vortices of the original fields \cite{ConfinementIV,ConfinementVIII,ConfinementXI,MCG,MCGFinding}. This allows one to investigate the significance of centre vortices through the projected links $Z_\mu(x)$.

The connection between centre vortices and confinement is apparent through space-time Wilson loops of size $R\times T$, which asymptotically give access to the static quark-antiquark potential $V(r)$,
\begin{align} \label{eq:staticquarkpotential}
    \langle W(R,T) \rangle \sim \exp\left(-V(r) \,a\,T\right) \,, && \! r=Ra\,, && \! T \text{ large} \,.
\end{align}
The percolation of centre vortices through spacetime implies an area law for the Wilson loop, Eq.~(\ref{eq:arealaw}) \cite{PvortexStructure,ConfinementVI,ConfinementVII,VortexAreaLawI}, which allows one to extract a potential $V(r)$ that linearly rises with the separation $r$. In the confined phase, this has been seen in numerical simulations for both SU(2) \cite{ConfinementI,ConfinementIV} and SU(3) \cite{ConfinementVIII,ConfinementXI}.

In contrast, centre vortices in the deconfined phase amount to a vanishing quark-antiquark potential \cite{ConfinementV}, signalled by a trivial expectation value for space-time Wilson loops. This is a natural consequence of vortices no longer percolating all four dimensions, indicating an inherent change to the bulk vortex structure as the critical temperature $T_c$ is crossed. This alludes to the phase transition being geometric in nature, as also realised by a construction of centre-\textit{electric} fluxes \cite{Geometric}. In SU(2), it is understood the vortex sheet shifts to principally align with the temporal dimension, though the structure still percolates in the spatial dimensions \cite{PvortexStructure,ConfinementV,ConfinementVI,ConfinementVII}. This results specifically in an absence of vortices piercing space-time Wilson loops, and a trivial expectation value in Eq.~(\ref{eq:staticquarkpotential}) follows. The same overarching change is expected in SU(3), and initial numerical results support this conclusion \cite{ConfinementXII,Branching}. We will elucidate this property by visualising the centre vortex structure below and above $T_c$ and quantifying the observations with new statistical measures, thereby affirming the geometric nature of the deconfinement phase transition in QCD.

In this work, we utilise five pure-gauge ensembles of 100 configurations, two below $T_c$ and three above $T_c$, each with a spatial volume of $32^3$ and fixed isotropic lattice spacing $a=0.1\,$fm. The ensembles are generated using Hybrid Monte Carlo \cite{HMCI,HMCII} with an Iwasaki renormalisation-group improved action \cite{IwasakiI,IwasakiII}. The temporal extents and corresponding temperatures are summarised in Table~\ref{tab:ensembledetails},
\begin{table}[b]
\caption{\label{tab:ensembledetails} The number of sites $N_t$ in the temporal dimension and corresponding temperatures, both in MeV and in terms of the critical temperature $T_c = 270\,$MeV, for each ensemble.}
\begin{ruledtabular}
\begin{tabular}{D{.}{.}{2.0}D{.}{.}{3.2}D{.}{.}{1.4}}
\multicolumn{1}{c}{$N_t$} & \multicolumn{1}{c}{$T$ (MeV)} & \multicolumn{1}{c}{$T/T_c$} \\
\colrule
12 & 164.4 & 0.609 \\
8 & 246.6 & 0.913 \\
6 & 328.8 & 1.218 \\
5 & 394.6 & 1.461 \\
4 & 493.3 & 1.827
\end{tabular}
\end{ruledtabular}
\end{table}
where we take $T_c = 270\,$MeV \cite{Tc}.

\section{Visualisations} \label{sec:visualisations}
We now move to qualitatively analyse the centre vortex structure at finite temperature, drawing on visualisation techniques previously established in Ref.~\cite{Visualisations}. The basic construction is reiterated briefly here. Vortices exist on the dual lattice. This allows the plaquette in the centre-projected field to be written as \cite{ConfinementXII,Branching}
\begin{equation} \label{eq:projectedplaquete}
    P_{\mu\nu}(x) = \exp\left(\frac{\pi i}{3} \,\epsilon_{\mu\nu\kappa\lambda} \,m_{\kappa\lambda}(\bar{x}) \right) \,,
\end{equation}
where $\bar{x} = x + \frac{a}{2}(\hat{\mu} + \hat{\nu} - \hat{\kappa} - \hat{\lambda})$ and $m_{\kappa\lambda}(\bar{x})\in\{-1,\,0,\,1\}$ defines the oriented centre charge of the plaquette, such that $m_{\kappa\lambda}(\bar{x}) = -m_{\lambda\kappa}(\bar{x})$.

\begin{figure}
    \centering
    \includegraphics[width=\linewidth]{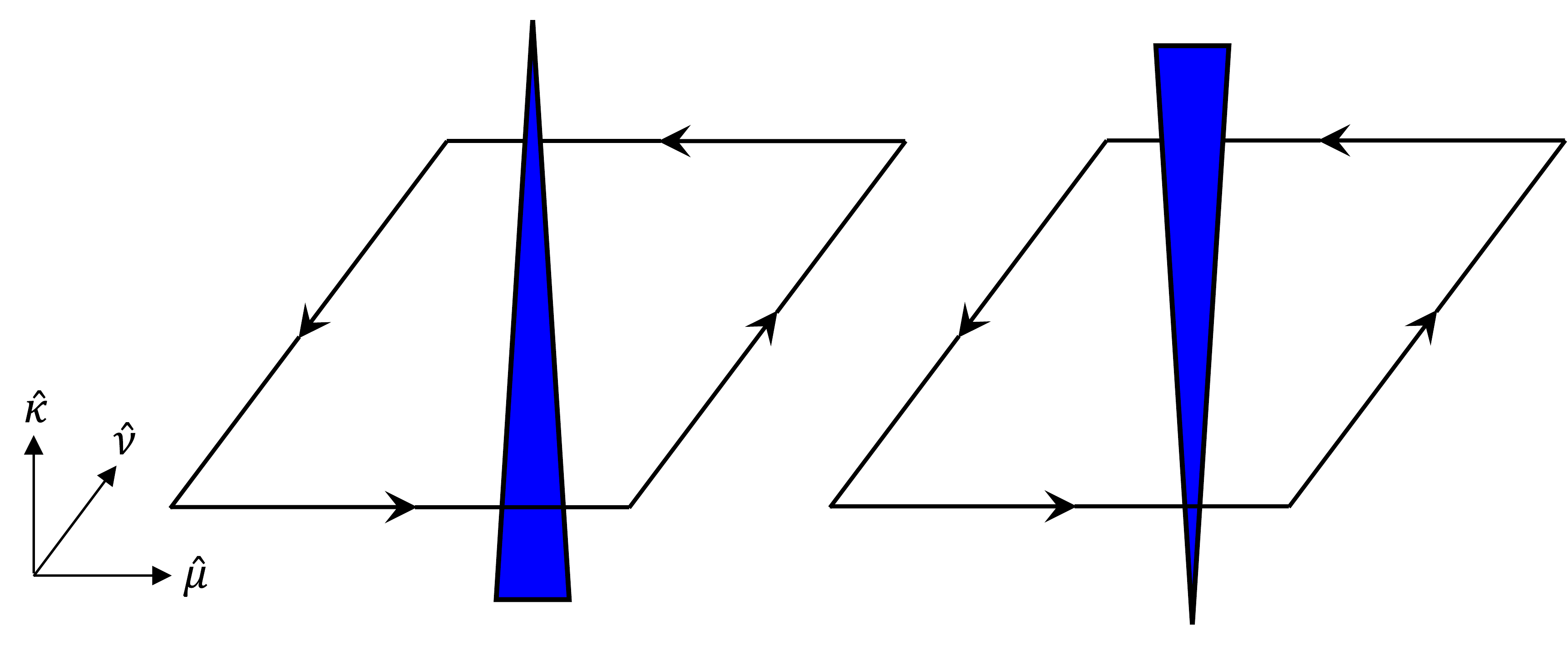}
    \caption{\label{fig:visconvention} The visualisation convention for centre vortices. An $m=+1$ vortex (\textbf{left}) is represented by a jet in the available orthogonal dimension, with the direction given by the right-hand rule. An $m=-1$ vortex (\textbf{right}) is rendered by a jet in the opposite direction.}
\end{figure}

To construct a 3D visualisation we slice over a given dimension, which corresponds to fixing the value of $\lambda$ in Eq.~(\ref{eq:projectedplaquete}). This leaves one orthogonal direction $\hat{\kappa}$ in the three-dimensional slice that can be used to identify the plaquette. For each nontrivial plaquette, the vortex is accordingly rendered as a jet pointing in the $m_{\kappa\lambda}(\bar{x})\,\hat{\kappa}$ direction and piercing the plaquette. Due to the presence of the Levi-Civita symbol in Eq.~(\ref{eq:projectedplaquete}), this effectively corresponds to implementing a right-hand rule for the orientation of the jet. As such, the visualisations exclusively show the flow of $m=+1$ centre charge. This convention is demonstrated in Fig.~\ref{fig:visconvention}.

\begin{figure*}
    \centering
    \includegraphics[width=0.48\textwidth]{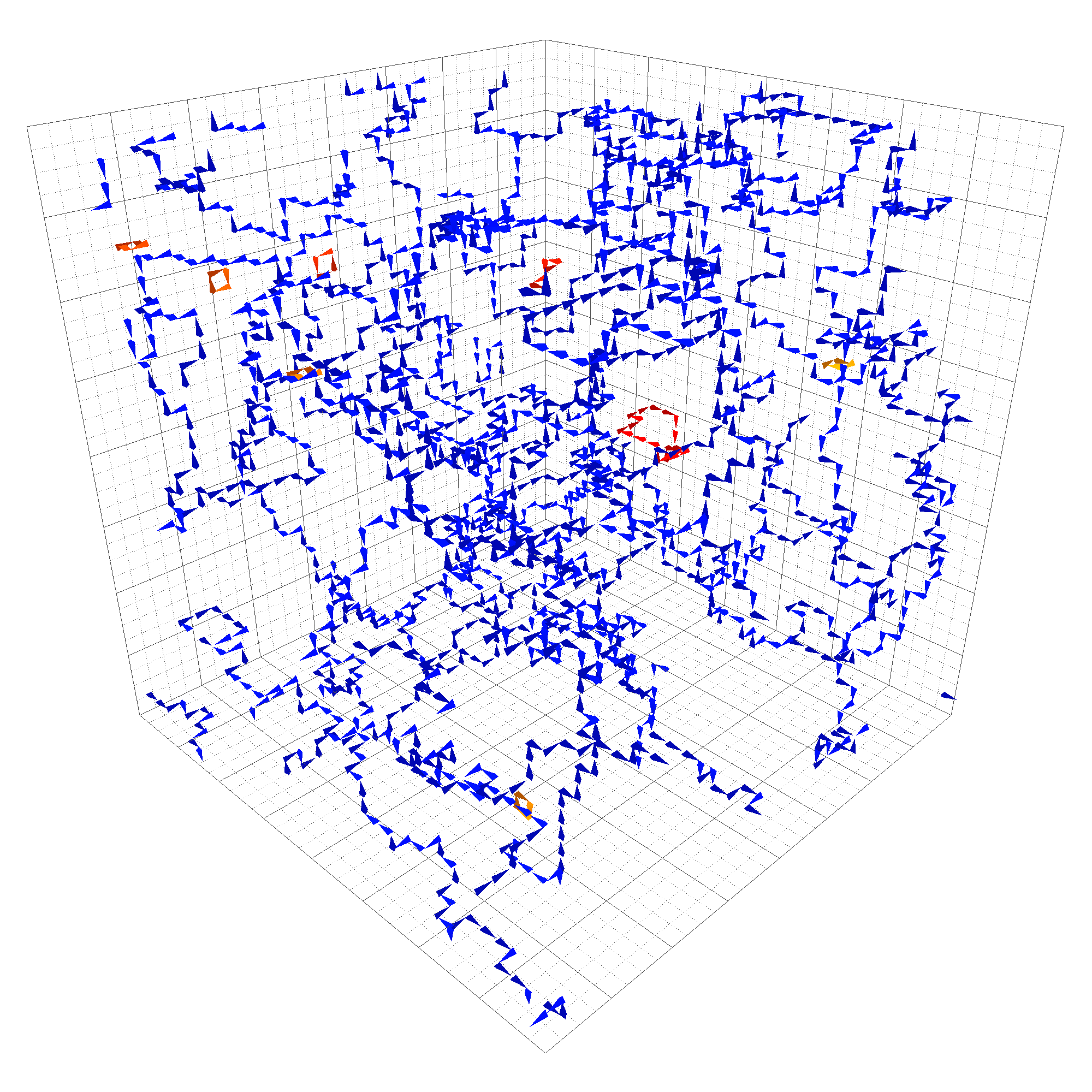}
    \includegraphics[width=0.47\textwidth]{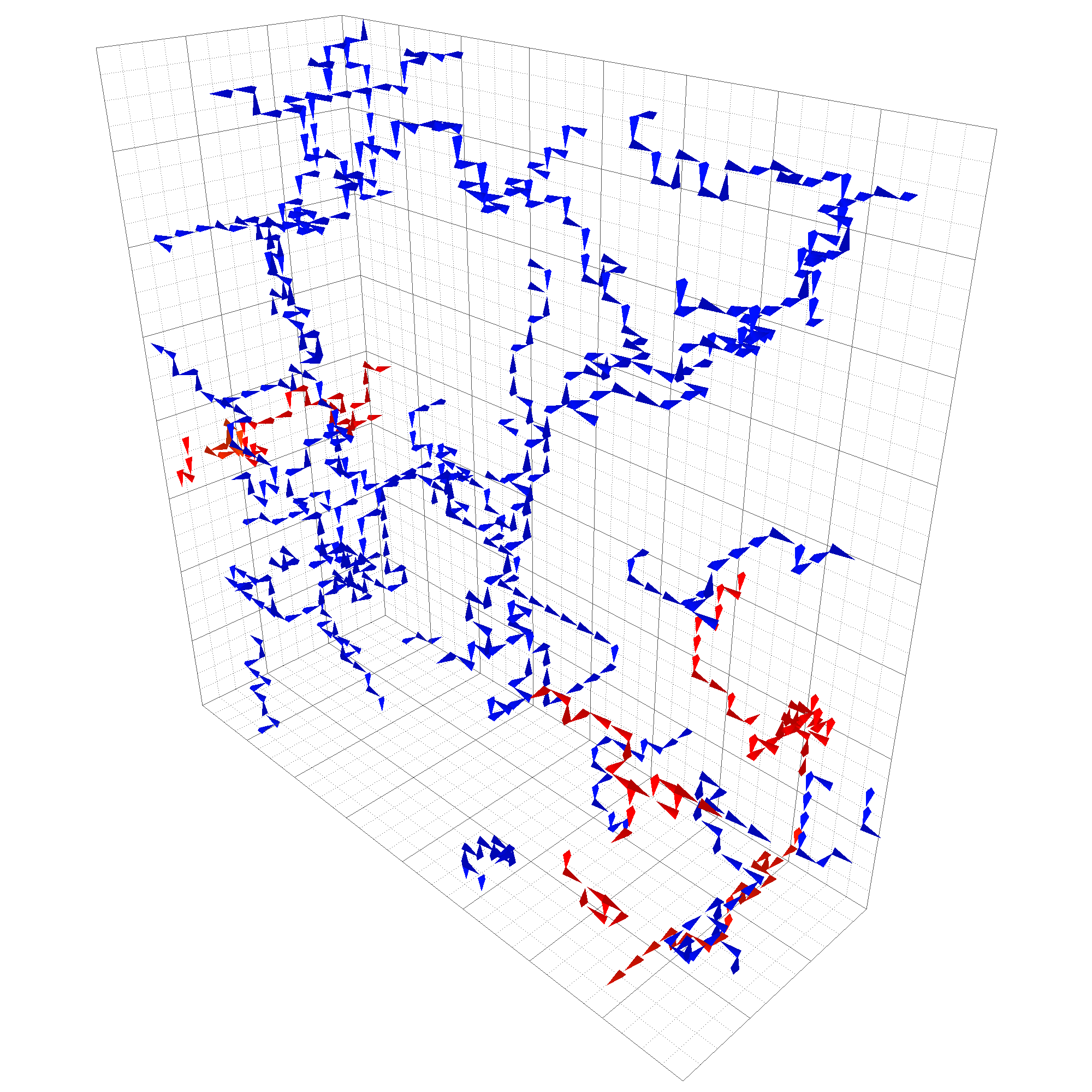}
    \vspace{-1em}
    \caption{\label{fig:Nt12Vis} Centre vortex structure in temporal slices (\textbf{left}) and spatial slices (\textbf{right}) below the critical temperature at $T=0.609\,T_c$. In this and the following four illustrations, the shorter dimension in the right-hand visualisation is the temporal direction.}
\end{figure*}

\begin{figure*}
    \centering
    \includegraphics[width=0.48\textwidth]{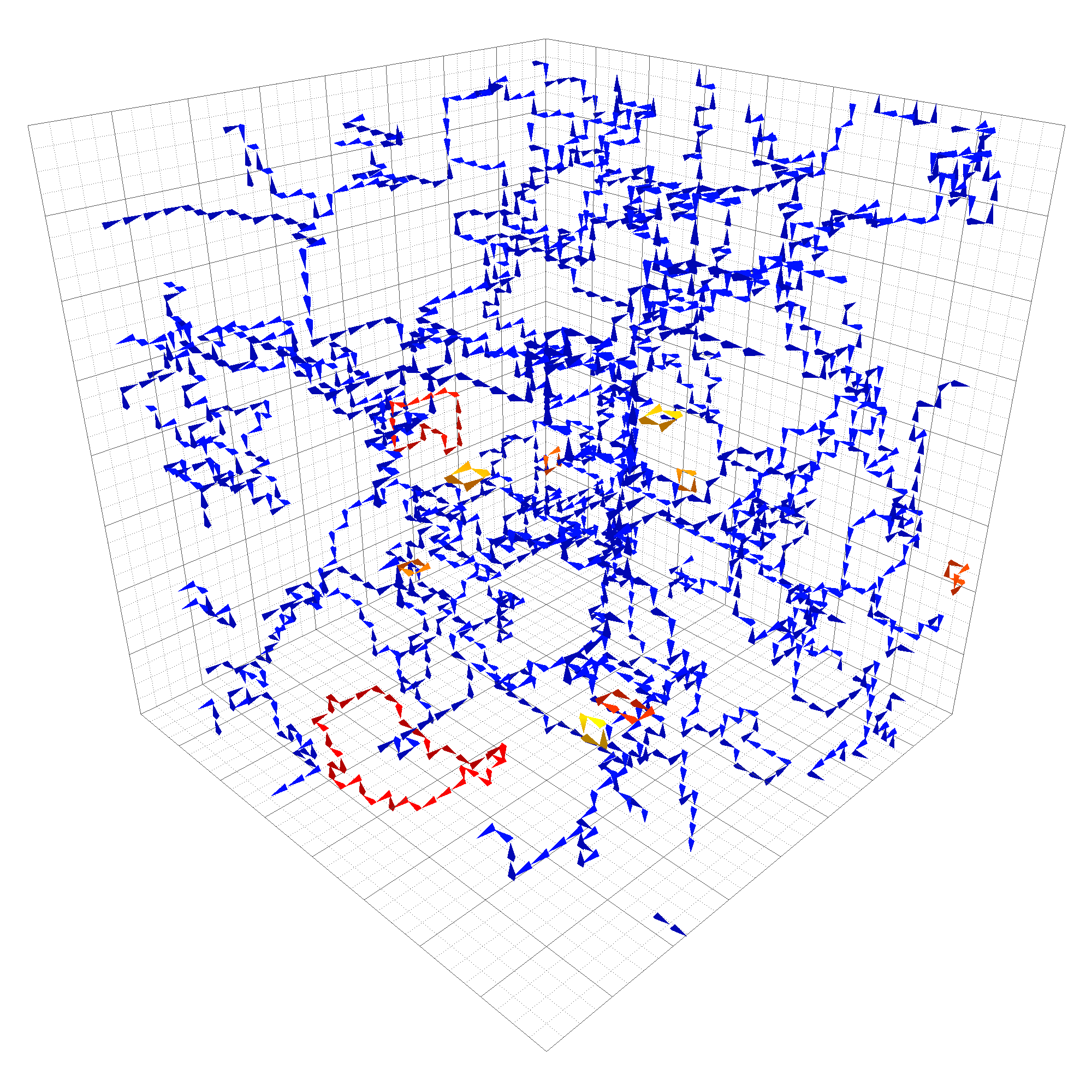}
    \includegraphics[width=0.47\textwidth]{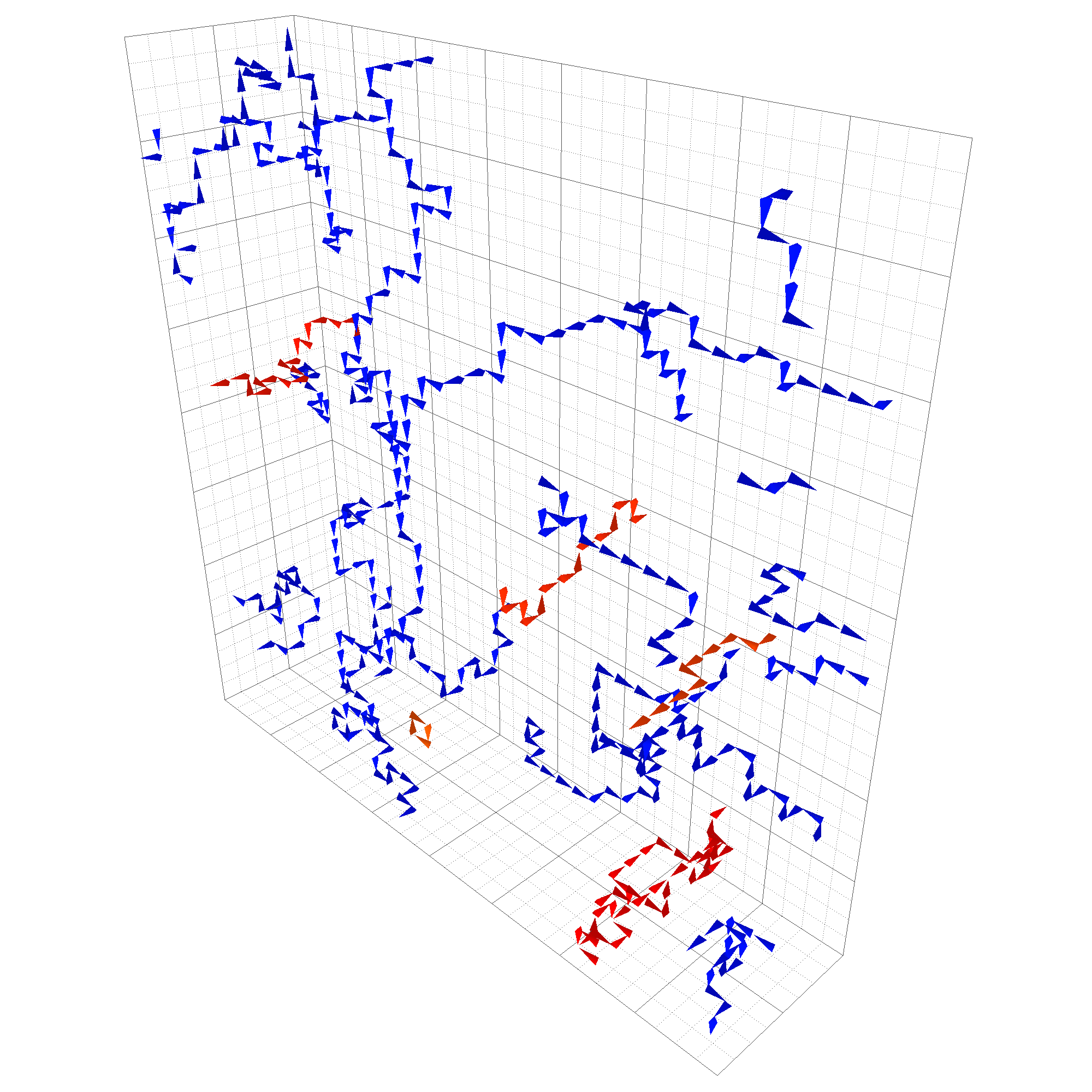}
    \vspace{-1em}
    \caption{\label{fig:Nt8Vis} Centre vortex structure in temporal slices (\textbf{left}) and spatial slices (\textbf{right}) below the critical temperature at $T=0.913\,T_c$.}
\end{figure*}

\begin{figure*}
    \centering
    \includegraphics[width=0.48\textwidth]{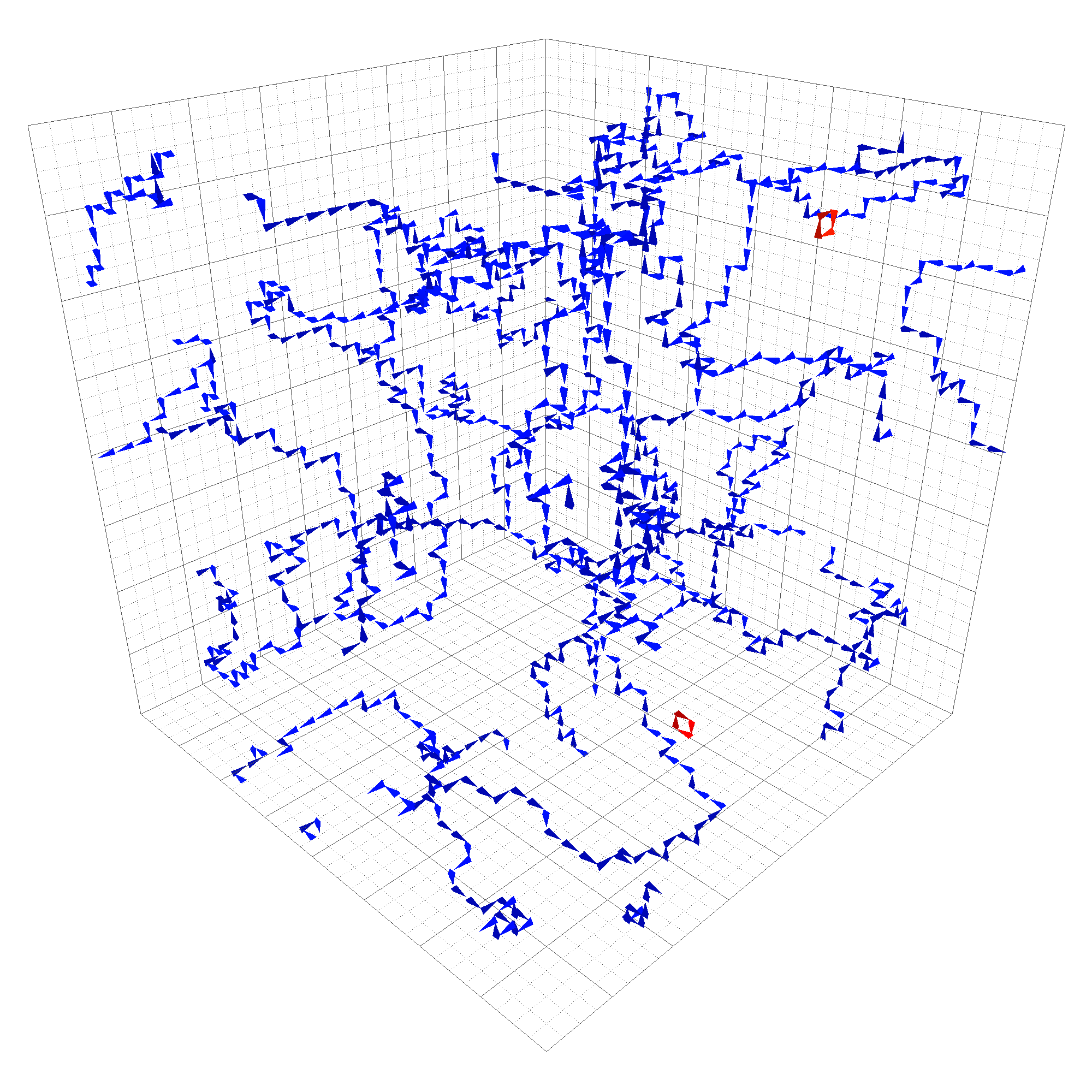}
    \includegraphics[width=0.47\textwidth]{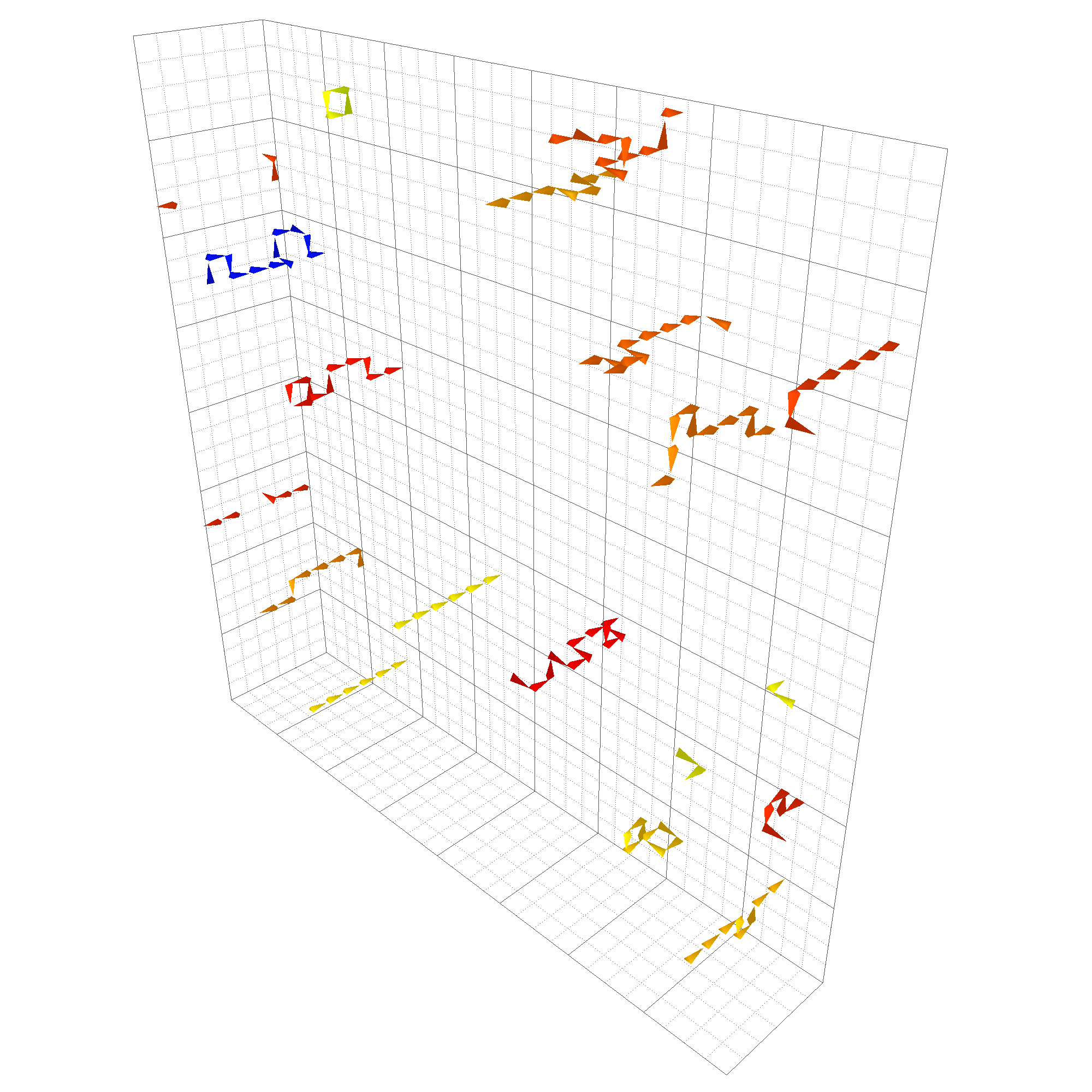}
    \vspace{-1em}
    \caption{\label{fig:Nt6Vis} Centre vortex structure in temporal slices (\textbf{left}) and spatial slices (\textbf{right}) above the critical temperature at $T=1.218\,T_c$.}
\end{figure*}

\begin{figure*}
    \centering
    \includegraphics[width=0.48\textwidth]{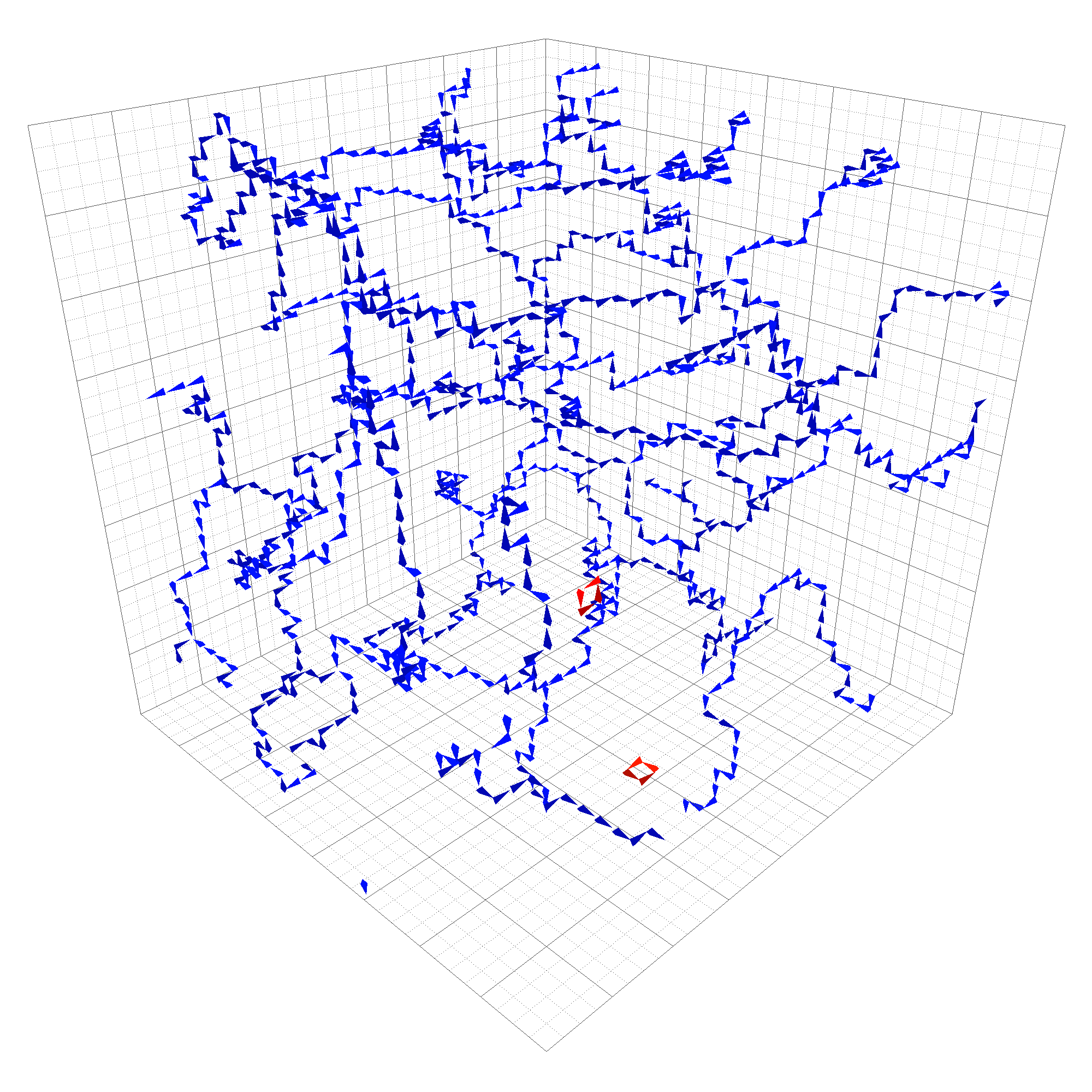}
    \includegraphics[width=0.47\textwidth]{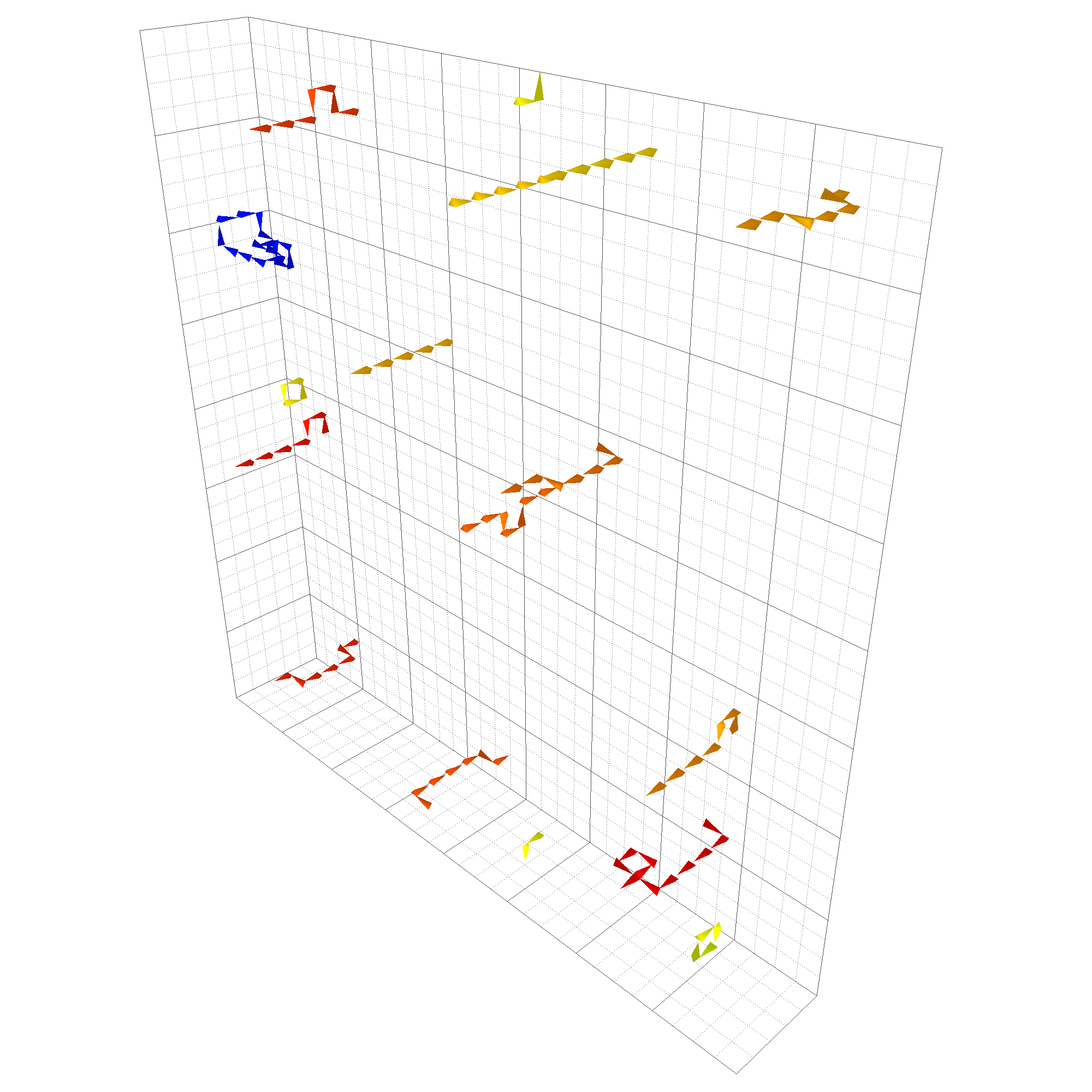}
    \vspace{-1em}
    \caption{\label{fig:Nt5Vis} Centre vortex structure in temporal slices (\textbf{left}) and spatial slices (\textbf{right}) above the critical temperature at $T=1.461\,T_c$.}
\end{figure*}

\begin{figure*}
    \centering
    \includegraphics[width=0.48\textwidth]{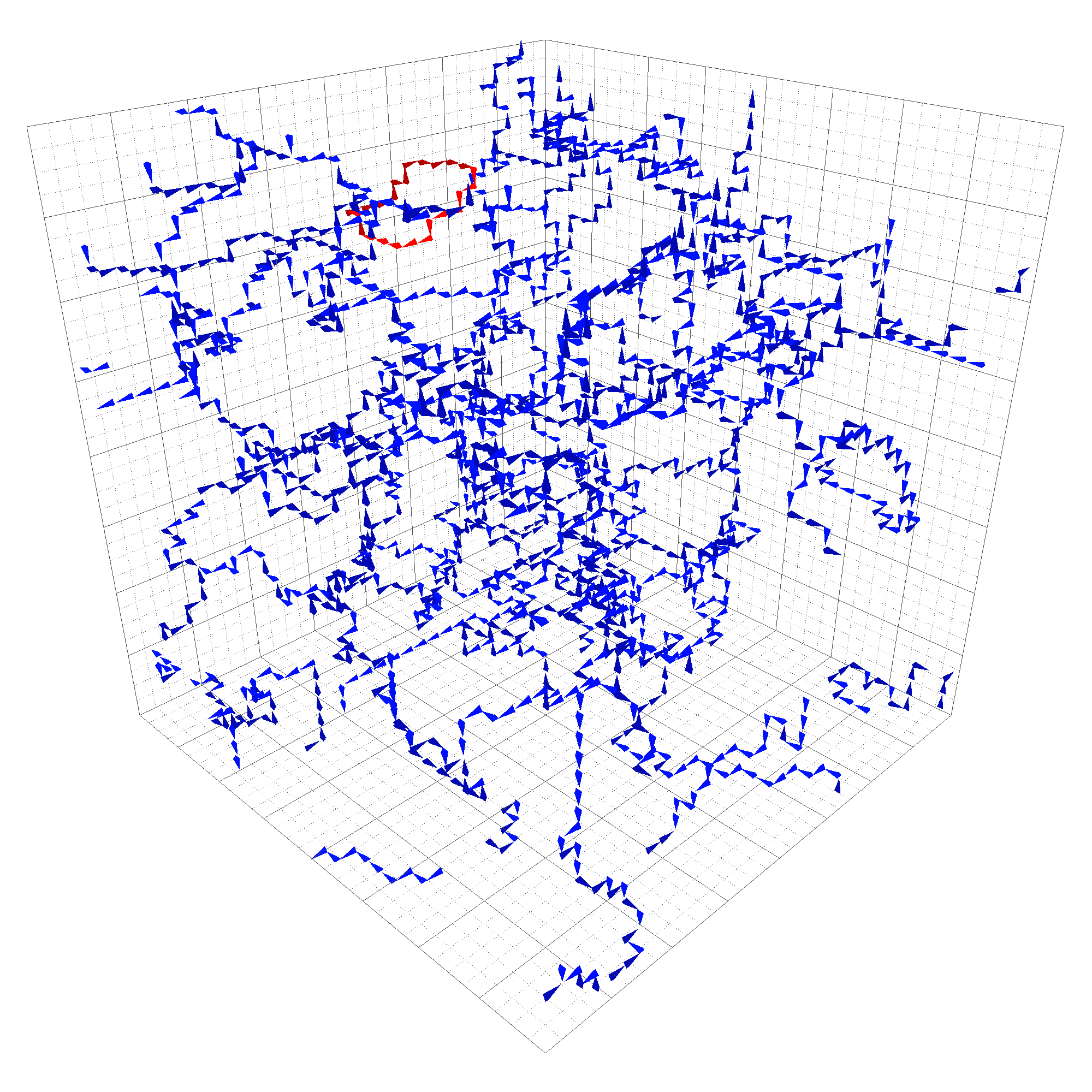}
    \includegraphics[width=0.47\textwidth]{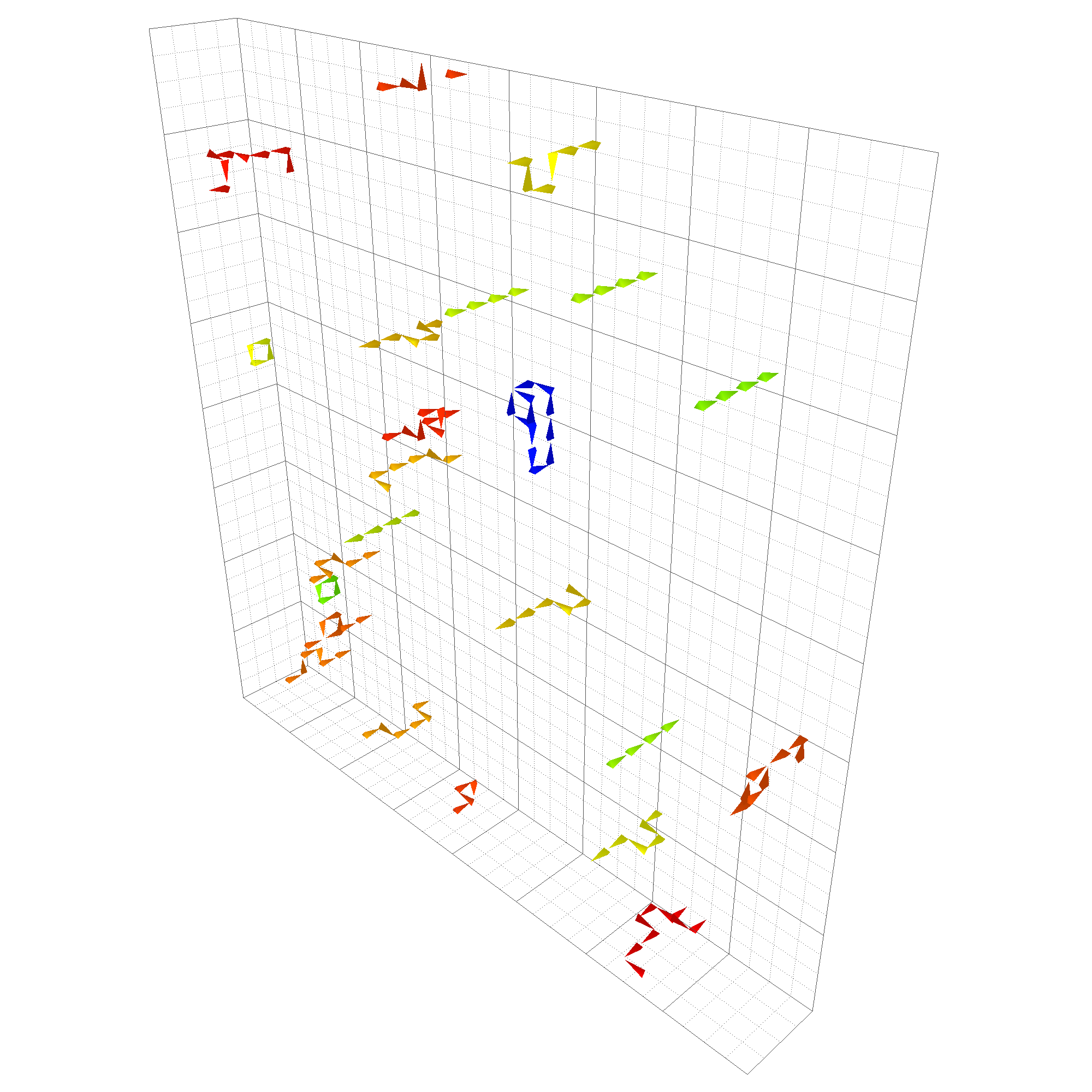}
    \vspace{-1em}
    \caption{\label{fig:Nt4Vis} Centre vortex structure in temporal slices (\textbf{left}) and spatial slices (\textbf{right}) above the critical temperature at $T=1.827\,T_c$.}
\end{figure*}

Utilising the SU(3) cluster identification algorithm developed in Ref.~\cite{StructureDynamical}, we present typical vortex structures in three-dimensional slices at increasing temperature in Figs.~\ref{fig:Nt12Vis}--\ref{fig:Nt4Vis}. We separately show the structure in temporal slices ($\lambda=4$) and spatial slices ($\lambda=1,\,2,\,3$), which is important to reveal the primary features of the phase transition. The structure in the spatial slices is insensitive to the choice of spatial dimension sliced over, and thus we present visualisations for the $x$ dimension ($\lambda=1$).

In the confined phase (Figs.~\ref{fig:Nt12Vis} and \ref{fig:Nt8Vis}), the visualisations reveal a single large percolating cluster (coloured in blue) that dominates the vortex structure in both temporal and spatial slices. A handful of smaller secondary clusters are scattered throughout the lattice. This is consistent with previous observations on near-zero temperature pure gauge configurations \cite{Visualisations,StructureDynamical}, indicating the external features of the vortex structure remain unchanged as we approach the critical temperature from below.

Moving above the critical temperature (Figs.~\ref{fig:Nt6Vis}--\ref{fig:Nt4Vis}) results in a shift in the behaviour of vortex matter. Clearly, the temporal and spatial structures diverge. The temporal slices are still dominated by a single large connected cluster, indicating the vortex structure remains percolating in the spatial dimensions. In contrast, the spatial slices instead reveal an abundance of small vortex clusters mostly parallel to the short temporal dimension, which are closed under periodic boundary conditions. This establishes that the vortex sheet in the deconfined phase principally aligns with the temporal dimension, matching the observation in SU(2). As a timelike surface, the vortex sheet rarely cuts through space-time plaquettes above $T_c$. This underlies the strong preference for vortices to pierce space-space plaquettes in spatial slices, which are oriented with the temporal axis.

Nonetheless, not all clusters wind around the temporal dimension, and there are still remnant fluctuations where a vortex line ``twists" to temporarily propagate in a spatial dimension. This latter effect could be in part due to ambiguity in the precise location of projected vortices within the physical ``thick" centre vortices \cite{ConfinementIV}. Still, the most significant feature of the spatial slices above the critical temperature is an absence of vortex lines that span the spatial extent and pierce opposite space-time faces, demonstrating a collapse in vortex geometry that results in the loss of percolation.

To elaborate on this, we provide animations over the temporal and spatial slices for each temperature throughout Figs.~\ref{fig:Nt12Animation}--\ref{fig:Nt4Animation} in the Supplemental Material. This aids in understanding the four-dimensional nature of the centre vortex sheet. Below $T_c$, the vortex structure changes considerably on a per-slice basis in both temporal and spatial slices, with the vortex lines moving erratically around the lattice. This simply indicates the vortex sheet is not aligned with any particular dimension, implying the observed three-dimensional structure depends on where one slices. The animations above $T_c$ are more intriguing. In temporal slices, the vortex structure is effectively ``frozen", only experiencing a slight shimmer or oscillation from slice to slice. This is yet another manifestation that the vortex sheet is oriented with the temporal dimension, as this signifies the spatial structure is predominantly unchanged regardless of where along the Euclidean time dimension one slices. On the other hand, animating over the spatial slices above $T_c$ shows there is still substantial change in the position of the short vortex lines. This is a consequence of the fact that the centre vortex sheet still percolates in the spatial dimensions.

For an analogy to the deconfined phase, one can imagine a three-dimensional cylinder aligned with the $z$-axis. If slicing along said $z$ dimension, one would find an identical circle regardless of where the slice is taken, which is the equivalent scenario to slicing the centre vortex structure through the temporal dimension. By contrast, slicing along the $x$ or $y$ dimensions would result in two disconnected lines parallel to the $z$-axis. Importantly, the position of and distance between these lines varies depending on the slice coordinate. This matches the arrangement of vortices in spatial slices, with many disconnected clusters winding around the temporal dimension but still moving over the lattice through each slice.

At this point, an essential consideration is whether the alignment of the vortex sheet is a genuine dynamical effect, or a byproduct of algorithmic issues. For instance, it could indicate the presence of non-ergodicity in the Markov chain of lattice QCD, with the persistent alignment observed throughout the ensemble due to topological locking. Another possibility is that the alignment arises specifically from the low number of temporal lattice sites required at high temperatures on coarser lattices. These concerns are especially prescient with the context that vortices are stiff \cite{ConfinementVII,ConfinementXII}, such that configurations in the deconfined phase wherein the vortex sheet is not primarily aligned with the short temporal dimension are suppressed. We investigate these possibilities with greater detail in Sec.~\ref{subsec:algissues}. Using a novel measure to quantify the alignment through the freezing of the vortex structure in temporal slices, we conclude the vortex sheet alignment with the temporal dimension is a genuine dynamical effect.

The fact temporal slices comprise a percolating cluster in both phases raises the question of whether any more subtle underlying changes occur to the vortex structure. Comparing Figs.~\ref{fig:Nt12Vis}--\ref{fig:Nt4Vis}, it is visually clear the primary cluster in the temporal slices experiences a significant reduction in density of vortex matter as the critical temperature is crossed. Subsequently, this low density appears to gradually increase as the temperature climbs away from $T_c$, with the cluster at our highest temperature nearly matching that found in the confined phase. There is also a plausible drop in the number of secondary clusters found in the temporal slices above $T_c$, though it is unclear whether this is representative of the entire ensemble or a consequence of the specific slices displayed. These statistics, along with several others, will be investigated quantitatively in the following section.

\section{Vortex statistics} \label{sec:statistics}
Next, we investigate a number of statistical quantities related to centre vortices to bring to light any additional structural changes not immediately apparent through visualisation. Several such properties, such as vortex and branching point densities, have previously been investigated across the phase transition in SU(3) in Ref.~\cite{Branching}, though here we extend this to a broader range of temperatures. We further analyse the secondary clusters present throughout the volume and compute a correlation of the vortex structure between various slices of the lattice. We defer our branching point analysis to Sec.~\ref{sec:branchingpoints} where we perform a thorough investigation into the geometry of branching points as a function of temperature.

In this section, we display an extra point at $T/T_c = 0.114$ on our plots using data from Ref.~\cite{StructureDynamical}, obtained from a $32^3\times 64$ ensemble of 200 configurations, but with otherwise identical properties to our finite-temperature ensembles. Statistical errors are calculated through the standard deviation of 100 bootstrap ensembles.

\subsection{Vortex structure correlation} \label{subsec:correlation}
The key feature of the phase transition relevant to centre vortices is the alignment of the vortex sheet with the temporal dimension. This manifests in time slices of the lattice as a vortex cluster that undergoes minimal changes between successive slices. For this reason, we seek to calculate a correlation of the vortex structure between temporal slices as a means to quantify this change and investigate the extent to which it continues to evolve above $T_c$.

\begin{figure}
	\centering
	\includegraphics[width=\linewidth]{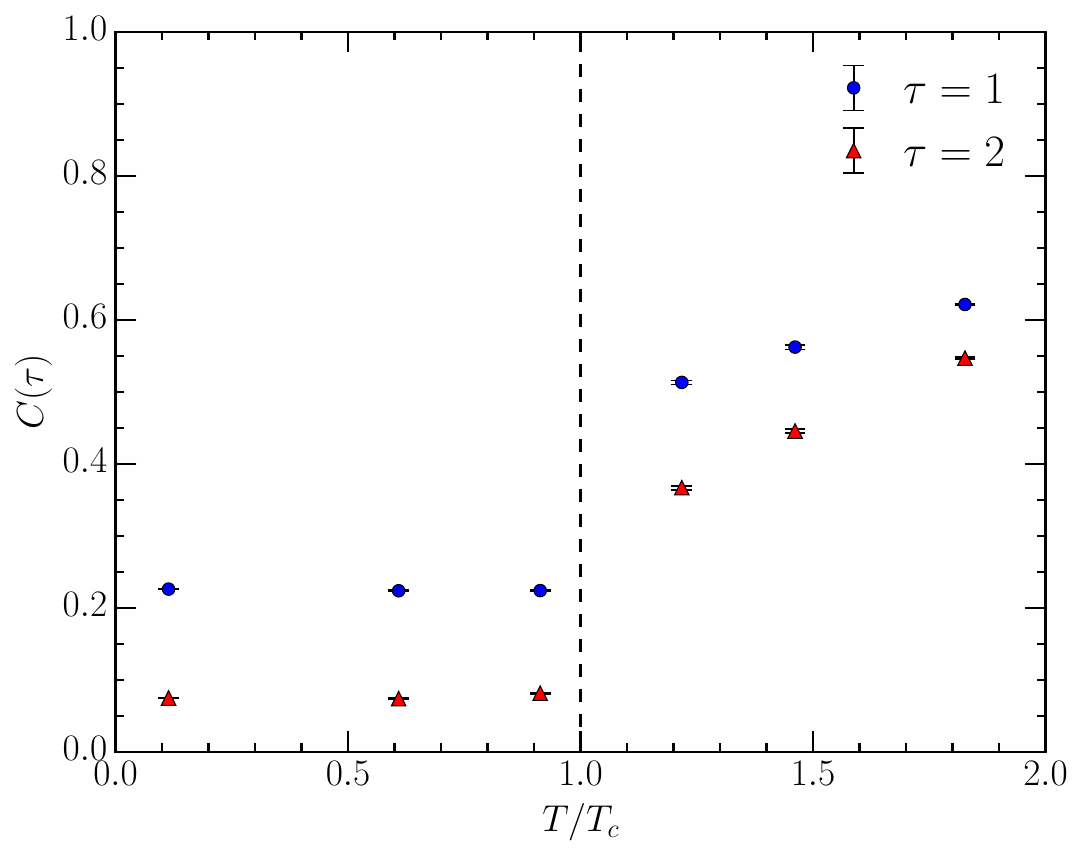}
	\caption{\label{fig:temporalcorrelation} The correlation $C(\tau)$ defined in Eq.~(\ref{eq:correlation}) for $\tau=1,\,2$ as a function of temperature. It is constant below $T_c$, though experiences a sudden jump at the phase transition where the vortices shift to principally align with the temporal dimension. Subsequently, it continues to increase above $T_c$.}
\end{figure}

This is achieved by first defining the indicator function,
\begin{equation}
	\chi_{ij}(\mathbf{x}, t; \tau) = 
	\begin{cases}
		1, & m_{ij}(\mathbf{x},t) \, m_{ij}(\mathbf{x},t+\tau) > 0 \\
		0, & \mathrm{otherwise}\\
	\end{cases} \,,
\end{equation}
which takes the value $1$ if a nontrivial plaquette at spatial position $\mathbf{x}$ in time slice $t$ is also pierced by a vortex in some later time slice $t+\tau$ with the same centre charge. As we are looking in temporal slices, here $i,\,j=1,\,2,\,3$. We can then define the correlation measure,
\begin{equation} \label{eq:correlation}
	C(\tau) = \frac{1}{N_\mathrm{vor}\,N_t} \,\sum_{\substack{\mathbf{x},\,t,\\i,\,j}} \chi_{ij}(\mathbf{x}, t; \tau) \,,
\end{equation}
where $N_\mathrm{vor}$ is the average number of vortices per temporal slice. This normalisation factor enforces the maximum value of $C(\tau)$ to $1$, corresponding to the case the vortex structure is completely unchanging along the temporal dimension. Due to periodic boundary conditions, $\tau$ is restricted to $\tau \leq \lfloor N_t/2 \rfloor$. Consequently, we compare $\tau=1,\,2$ across all our ensembles. This evolution is presented in Fig.~\ref{fig:temporalcorrelation}.

\begin{figure}
	\centering
	\includegraphics[width=0.8\linewidth]{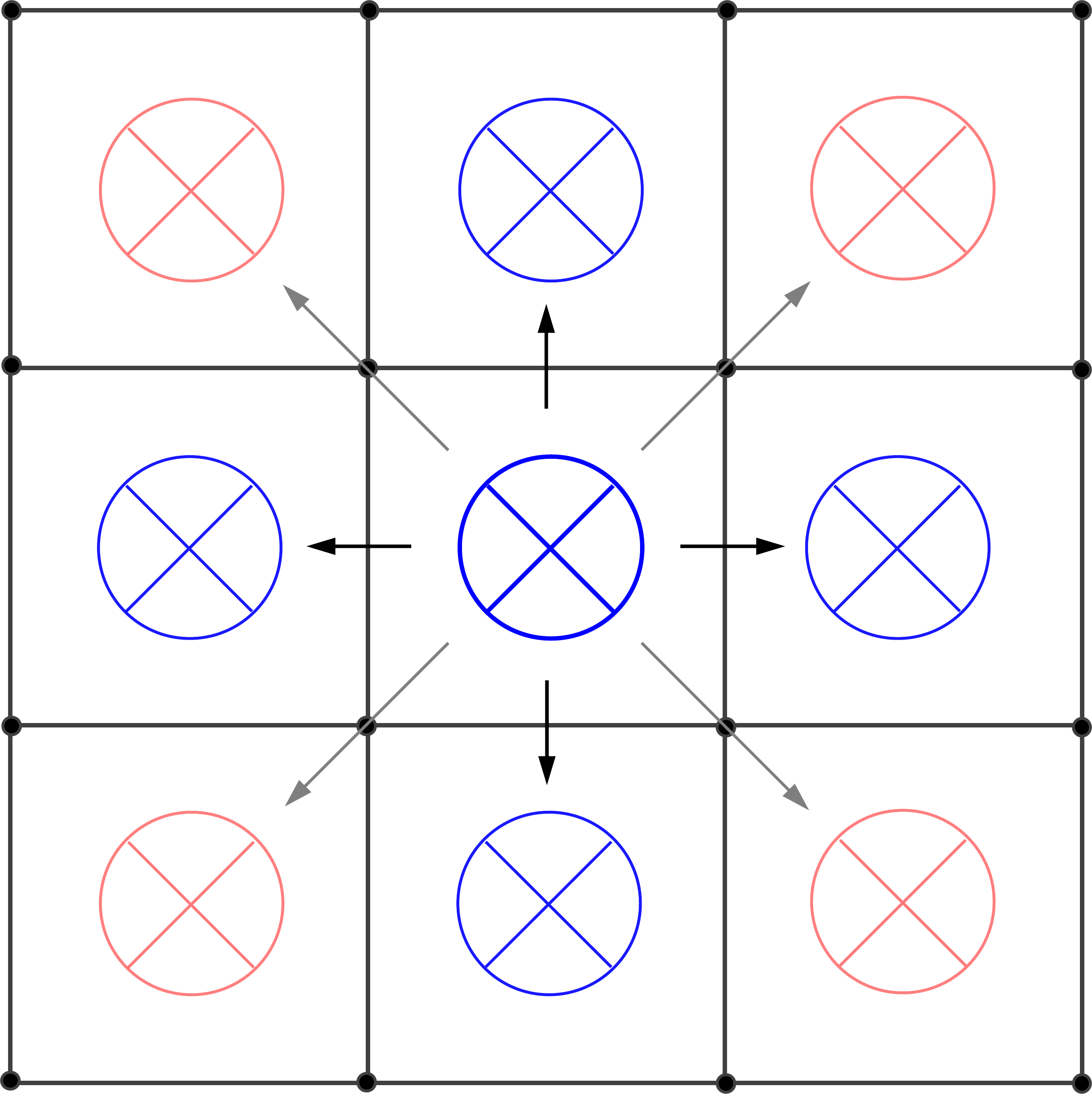}
	\caption{\label{fig:correlation} An illustration of the mechanism underlying vortex correlations. Initially, the centre plaquette is pierced by a vortex on some time slice. Provided there is some `smoothness' to the vortex structure, on the subsequent time slice it can either remain invariant or move to one of the surrounding plaquettes. In the simplest case, the vortex remains invariant or moves to one of the four directly adjacent plaquettes (dark arrows), corresponding to a probability of $1/5$ that the centre plaquette remains pierced on consecutive time slices. However, depending on the angle at which the vortex sheet passes through the time slice, the jet can also move to one of the four `diagonal' plaquettes (grey arrows). This implies a probability for the vortex to remain invariant of less than $1/5$.}
\end{figure}

Below $T_c$, the correlation is found to be constant with $C(1)$ slightly above $0.2$, whilst $C(2)$ unsurprisingly attains a substantially smaller value. It is natural to inquire as to whether one can infer any significance to these values. To do so, consider a slowly varying field such that a vortex sheet passes through two consecutive (three-dimensional) time slices with spatial positions in the neighbourhood of each other. Then, it is reasonable to assume that a pierced plaquette could either remain invariant or move to an adjacent plaquette. In the case that an equal probability is applied to each, there is a $1/5$ chance that a vortex line piercing a given space-space plaquette proceeds to pierce the same plaquette one time step later, thus producing a contribution to Eq.~(\ref{eq:correlation}). More generally, it is possible for the vortex sheet to pass through the three-dimensional slice at a sufficiently sharp angle that results in a jet on one slice moving a distance greater than the adjacent plaquette \cite{Visualisations}. For example, it could additionally move to one of the neighbouring diagonal plaquettes. Accounting for this possibility entails a probability smaller than $1/5$ for the vortex to remain invariant. This idea is illustrated in Fig.~\ref{fig:correlation}.

However, we nonetheless find $C(1)>1/5$, which can be attributed to a degree of ``smoothness" arising from the physical interpretation of the system. As the lattice spacing $a$ is decreased, $C(\tau)$ should increase (for fixed $\tau$) in accordance with the smaller physical distance between consecutive temporal slices. It is therefore fascinating our value of $C(1)$ lies slightly above $1/5$, indicating that a pierced plaquette typically tends to remain invariant or pierce one of the four adjacent plaquettes in consecutive slices. This physical smoothness in the vortex sheet can be seen in the supplementary animations (Figs.~\ref{fig:Nt12Animation}--\ref{fig:Nt4Animation}), where the movement of the vortex lines can to an extent be ``tracked" between successive slices.

As the phase transition is crossed, Fig.~\ref{fig:temporalcorrelation} shows $C(\tau)$ undergo a jump as the vortex structure predominantly aligns with the temporal dimension. The correlation continues to grow as the temperature increases above $T_c$, signalling the alignment becomes stronger at higher $T$. It is as yet unclear whether $C(\tau)$ would eventually reach a value of $1$ or instead plateau at some value $<1$. Exploring the extent of the alignment at even higher temperatures $T>2\,T_c$ could be the subject of future work. Furthermore, the separation between $C(1)$ and $C(2)$ lessens as the temperature rises. This is expected since $C(\tau)$ necessarily becomes constant with $\tau$ (and equal to $1$) if the vortex structure were completely frozen in the temporal dimension.

\subsection{Investigation of algorithmic issues} \label{subsec:algissues}
Before proceeding, as promised we will utilise the correlation measure as a means to investigate the possibility that the vortex sheet alignment is attributed to an algorithmic effect. For this we consider two avenues. First, to test the ergodicity of the Markov chain we thermalise 100 independent hot starts on two $T>T_c$ ensembles with $N_t=6$ and $N_t=4$. If the alignment remains a consistent feature throughout these ensembles, then we can be assured it is not a result of any non-ergodicity given each configuration started from an independent, completely random gauge field.

To explore a possible dependence on the gauge action, we perform this analysis using the standard Wilson action \cite{ConfinementAreaLawI}. Therefore, to proffer a fair comparison we generate an additional zero-temperature ensemble on the account that different actions may produce slight differences in the precise quantitative behaviour of vortex matter. The details for these new ensembles are provided in Table~\ref{tab:alignmentensembles}, along with the value of the correlation $C(\tau=1)$ on each.
\begin{table}[b]
	\caption{\label{tab:alignmentensembles} The $\beta$ value, lattice spacing $a$, spatial and temporal extents $N_s,\,N_t$ and the corresponding temperature for each ensemble used to investigate the alignment of the vortex sheet. The ensemble average of the correlation on consecutive slices $C(\tau=1)$ is also given. Simulations are performed with the Wilson action. The scale is set using Sommer scale data from Ref.~\cite{sommerscale}, where we take $r_0\simeq 0.5\,$fm.}
	\begin{ruledtabular}
		\begin{tabular}{cccD{.}{.}{2.0}cD{.}{.}{1.8}}
			\multicolumn{1}{c}{$\beta$} & \multicolumn{1}{c}{$a$ (fm)} & \multicolumn{1}{c}{$N_s$} & \multicolumn{1}{c}{$N_t$} & \multicolumn{1}{c}{$T/T_c$} & \multicolumn{1}{c}{$C(\tau=1)$}\\
			\colrule \\[-0.8em]
			\multirow{3}{*}{5.96} & \multirow{3}{*}{0.10} & \multirow{3}{*}{32} & 64 & 0.114 & 0.1954(2) \\
			& & & 6 & 1.218 & 0.4027(11) \\
			& & & 4 & 1.827 & 0.4989(16) \\[0.1em]
			\colrule \\[-0.8em]
			\multirow{2}{*}{6.42} & \multirow{2}{*}{0.05} & \multirow{2}{*}{64} & 12 & 1.218 & 0.4675(9) \\
			& & & 8 & 1.827 & 0.5182(4)
		\end{tabular}
	\end{ruledtabular}
\end{table}
We note that the value on the low-temperature ensemble of $C(1)\approx 0.195$ is slightly less than that from Fig.~\ref{fig:temporalcorrelation}, indicating that the vortex matter recovered from the Wilson action tends to be ``rougher" compared to the Iwasaki action of Refs.~\cite{IwasakiII,IwasakiI}. For this reason, we are content to accept smaller values of $C(1)$ at high temperature when compared against Fig.~\ref{fig:temporalcorrelation}, though there should still be a consistent significant increase from zero temperature across the ensemble.

For $N_t=6$ the correlation (in the ensemble average) has jumped from $C(1)\approx 0.195$ to $C(1)\approx 0.403$. The smallest value on any one configuration is $\approx 0.370$, indicating the strong alignment is present in each thermalised hot start. This is also represented by the small statistical uncertainty seen in Table~\ref{tab:alignmentensembles}. The same pattern is seen with $N_t=4$, on which the ensemble average has climbed to $C(1)\approx 0.499$. In this ensemble, there is one particularly interesting case. A single $N_t=4$ configuration exhibited a correlation value of $C(1)\approx 0.373$, a clear outlier compared to the next-smallest value of $C(1)\approx 0.479$. Although still displaying a preferential alignment with the temporal dimension compared to zero temperature, this configuration also featured two disconnected vortex lines winding around the long spatial dimensions. These inevitably pierce many additional space-time plaquettes. Such a configuration would likely not be obtained within a continuous Markov chain, highlighting the importance of thermalising independent hot starts. That said, this was the only configuration from the set with such behaviour. From this we conclude that the alignment of the vortex sheet as represented by the freezing in temporal slices is a genuine dynamical effect. It is not the upshot of non-ergodicity in the Markov chain in the original set of ensembles.

The second avenue explored in this regard is the scaling of the alignment in taking the continuum limit at fixed temperature,
\begin{align}
	a\to 0\,, && N_t\to\infty\,, && T=(aN_t)^{-1} \text{ fixed} \,.
\end{align}
For this purpose we generate two ensembles, again comprising 100 configurations, with half the lattice spacing as previously (i.e. $a=0.05\,$fm) but twice the lattice extent in all four dimensions. These accordingly have $N_t=12$ and $N_t=8$ for a direct comparison. The full details are also provided in Table~\ref{tab:alignmentensembles}. By increasing the number of lattice sites in the temporal dimension, there are more chances for the vortex sheet to curve. We are concerned this could result in a softening of the alignment in the continuum limit, and hence it is critical to establish that the alignment persists to a similar degree after substantially diminishing the lattice spacing.

In fact, Table~\ref{tab:alignmentensembles} reveals that the correlation has only increased in reducing $a$, in line with the discussion in the previous section regarding the physical ``smoothness" of the vortex sheet. One might still be sceptical that the increase in value of $C(1)$ is mild compared to the factor of $1/2$ reduction in lattice spacing. However, it is important to bear in mind that since the lattice spacing has been decreased isotropically, the distance a jet must move to pierce an adjacent space-space plaquette, in relation to Fig.~\ref{fig:correlation}, is also physically smaller. These are competing effects which appear to approximately cancel each other out, resulting in the observed small increase in correlation between consecutive temporal slices. We therefore believe this indicates the overall extent of the alignment is approximately the same across our coarse and fine ensembles examined herein. If one were to instead utilise an anisotropic lattice, where exclusively the temporal spacing is decreased, we would expect this to effect a more substantial increase in the value of $C(\tau=1)$.

\begin{figure*}
	\centering
	\includegraphics[width=0.47\textwidth]{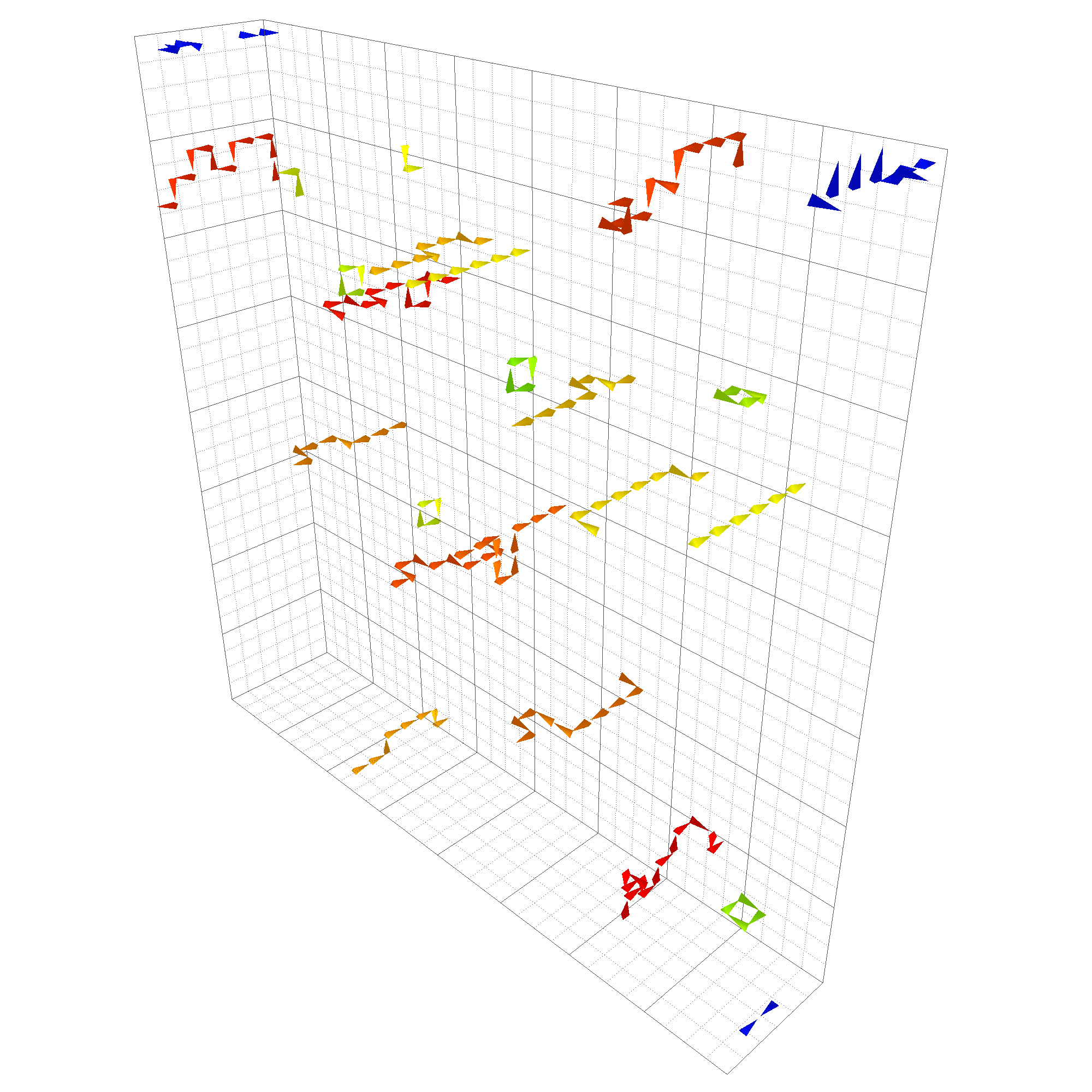}
	\includegraphics[width=0.47\textwidth]{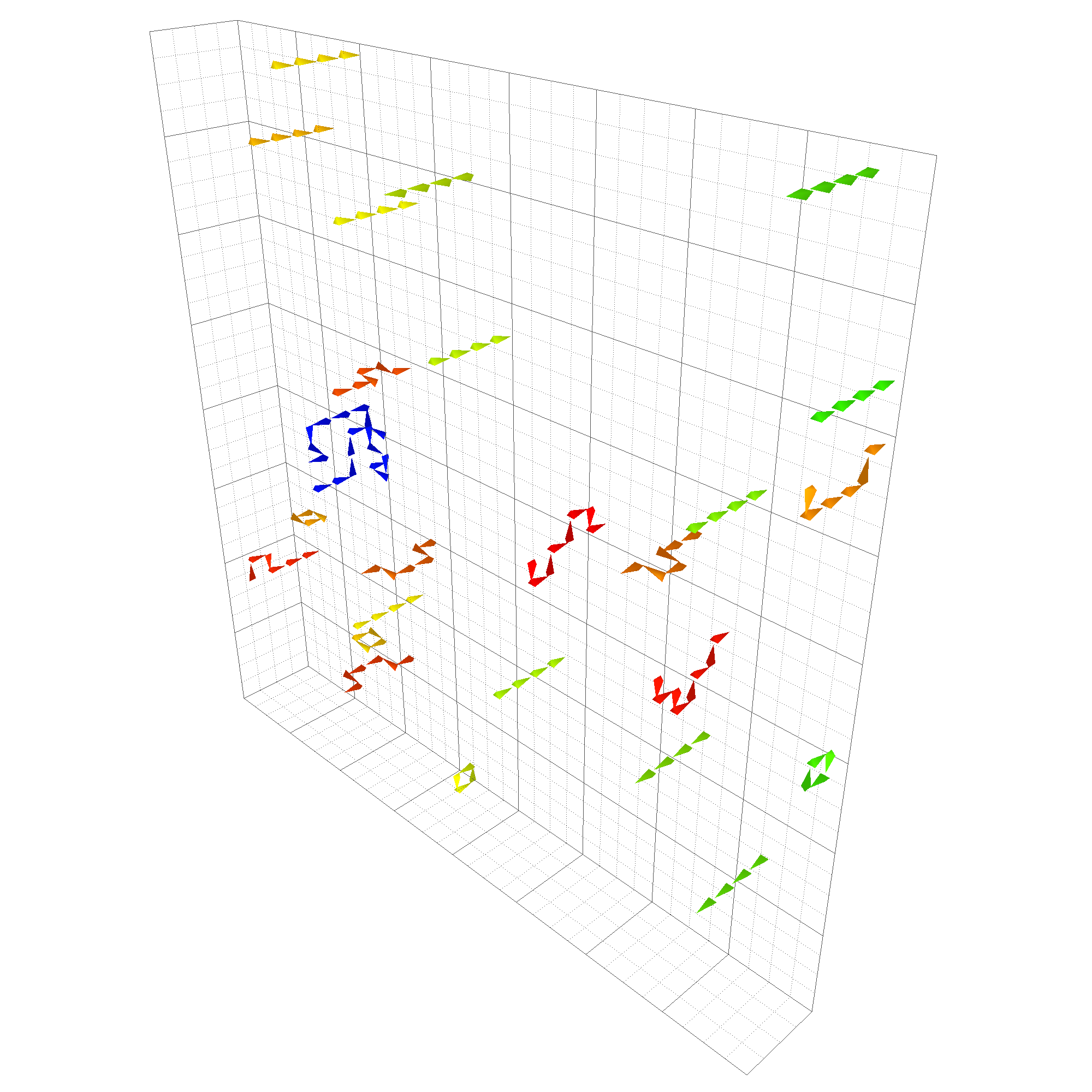}
	
	\vspace{1.2em}
	
	\includegraphics[width=0.47\textwidth]{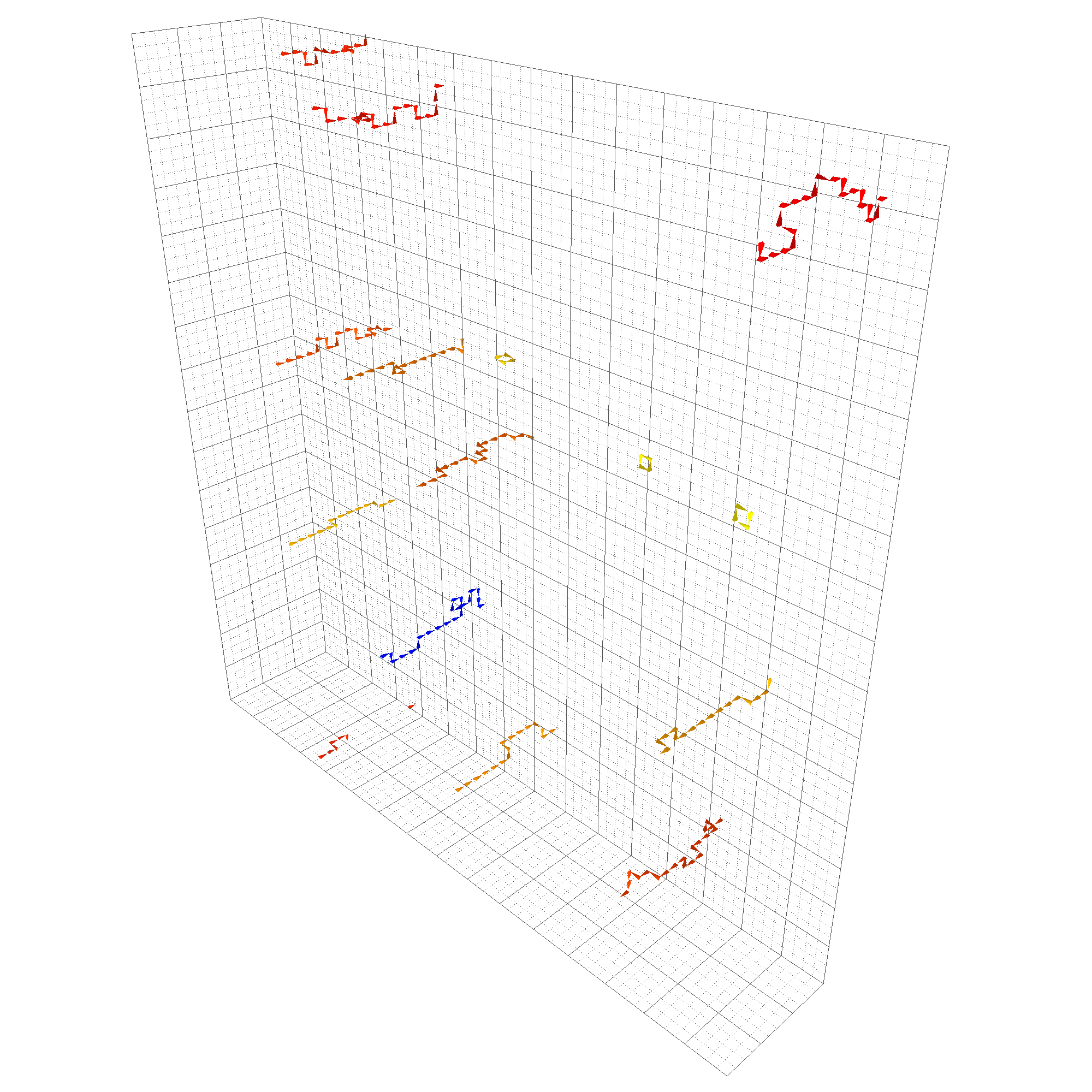}
	\includegraphics[width=0.47\textwidth]{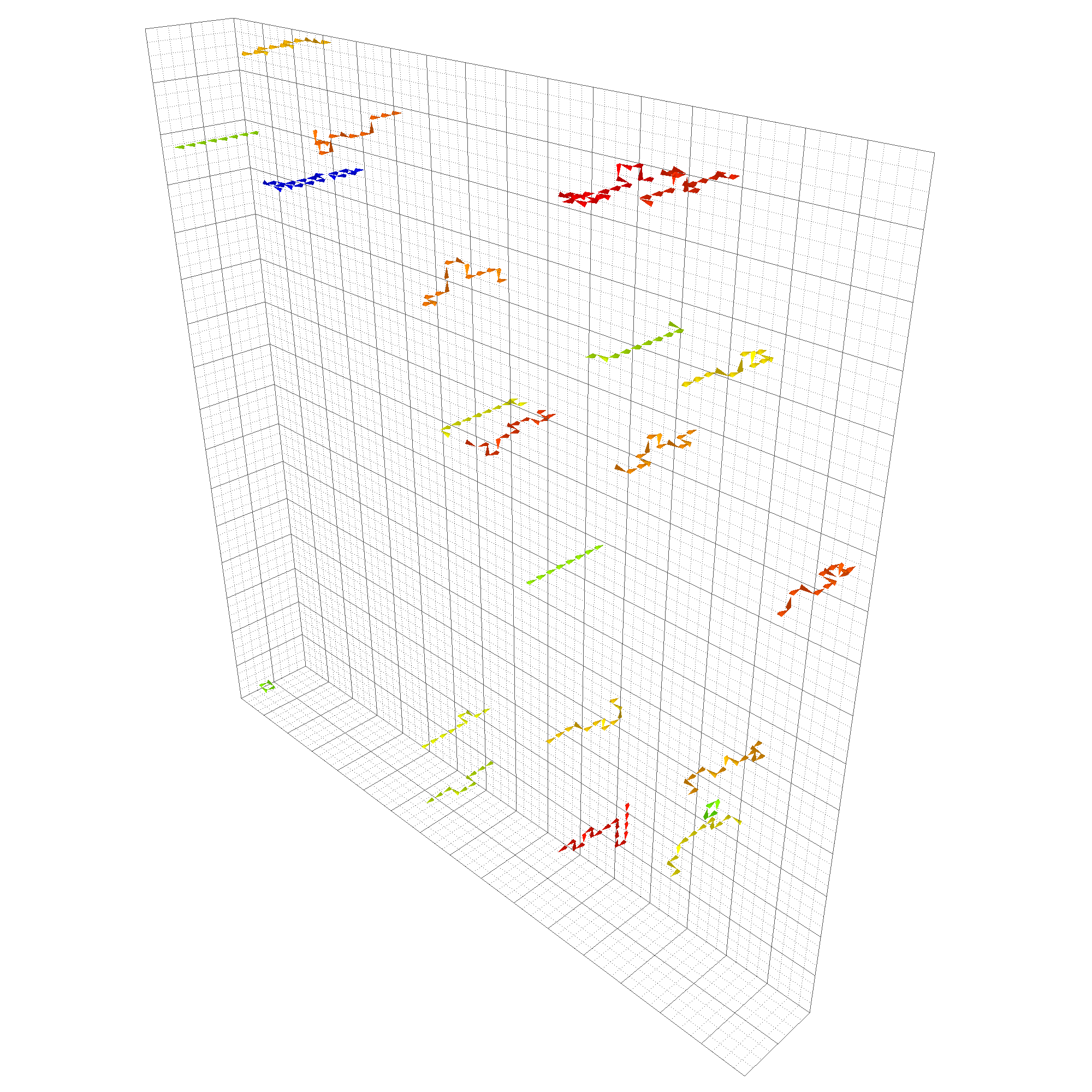}
	\caption{\label{fig:alignment} Centre vortex structure in a single spatial slice for each ensemble used to investigate the vortex sheet alignment with the temporal dimension: $\beta=5.96$ (\textbf{top}) and $\beta=6.42$ (\textbf{bottom}), with $T/T_c=1.218$ (\textbf{left}) and $T/T_c=1.827$ (\textbf{right}).}
\end{figure*}

We believe that this finding, in conjunction with the 100 independent hot starts, establishes the alignment and accompanying freezing in temporal slices as a physical effect rather than an algorithmic artefact. As a reference, in Fig.~\ref{fig:alignment} we provide visualisations of a typical spatial slice from each of the four additional high-temperature ensembles analysed in this section. Recall that the alignment manifests clearly in spatial slices as comprising primarily vortex lines which wind around the temporal axis. This is still seen to be true in each case here.

\subsection{Vortex density} \label{subsec:vortexdensity}
We now move to focus on the vortex area density, defined as the proportion of plaquettes pierced by a vortex. For a given three-dimensional lattice slice, this can be represented as
\begin{equation} \label{eq:vortexdensity}
    \hat{\rho}_\mathrm{vortex} = \frac{\text{Number of nontrivial plaquettes in slice}}{3\,V_\mathrm{slice}} \,,
\end{equation}
where $V_\mathrm{slice}$ is the number of lattice sites in the slice, and $\binom{3}{2} = 3$ counts the number of plaquettes at each site. We can then average over all slices along a given dimension. As defined in Eq.~(\ref{eq:vortexdensity}), $\hat{\rho}_\mathrm{vortex}$ is a dimensionless quantity, though can be converted to a physical quantity $\rho_\mathrm{vortex}$ by dividing by $a^2$, which gives proper scaling in the continuum limit \cite{ConfinementIII,ConfinementIV,ConfinementXI}.

\begin{figure}
    \centering
    \includegraphics[width=\linewidth]{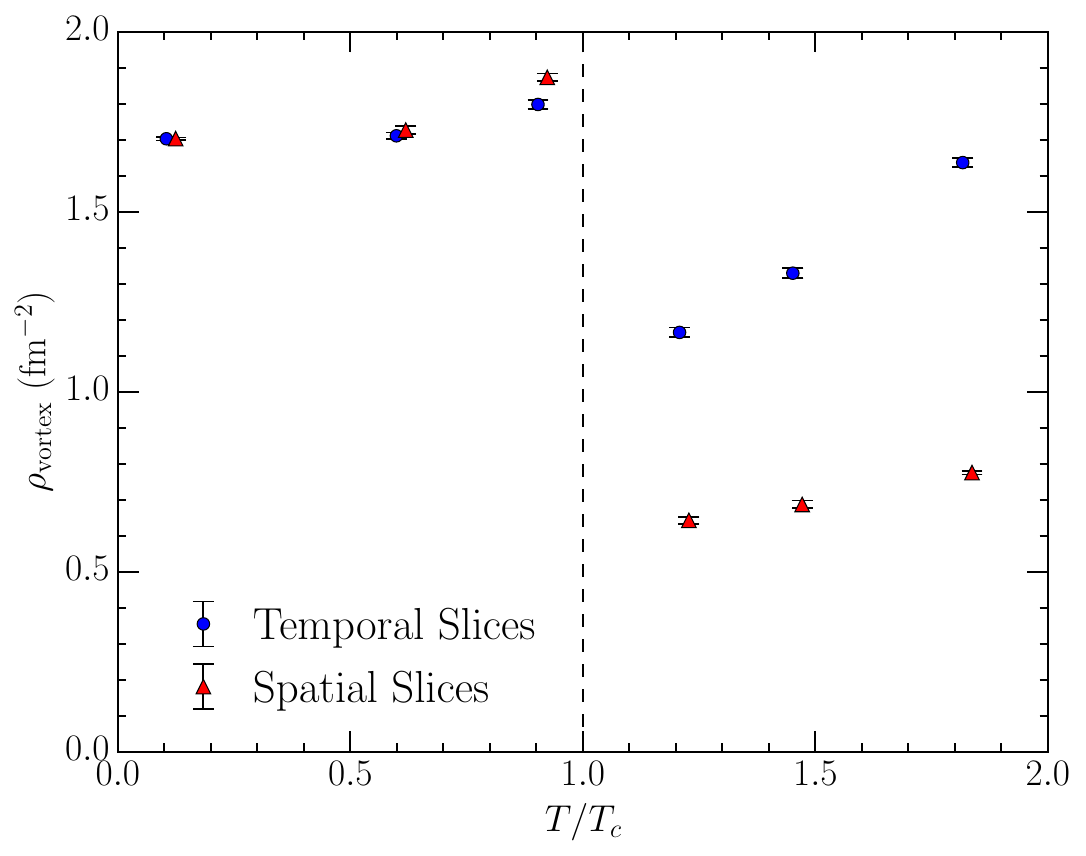}
    \caption{\label{fig:vortexdensity} The vortex area density in temporal and spatial slices of the lattice for each finite-temperature ensemble. The spatial slices experience a significant drop in vortex density at the phase transition, matching the absence of a percolating cluster. There is also an initial drop in the temporal slices, after which it subsequently increases with temperature.}
    
    \vspace{2em}
    
    \includegraphics[width=\linewidth]{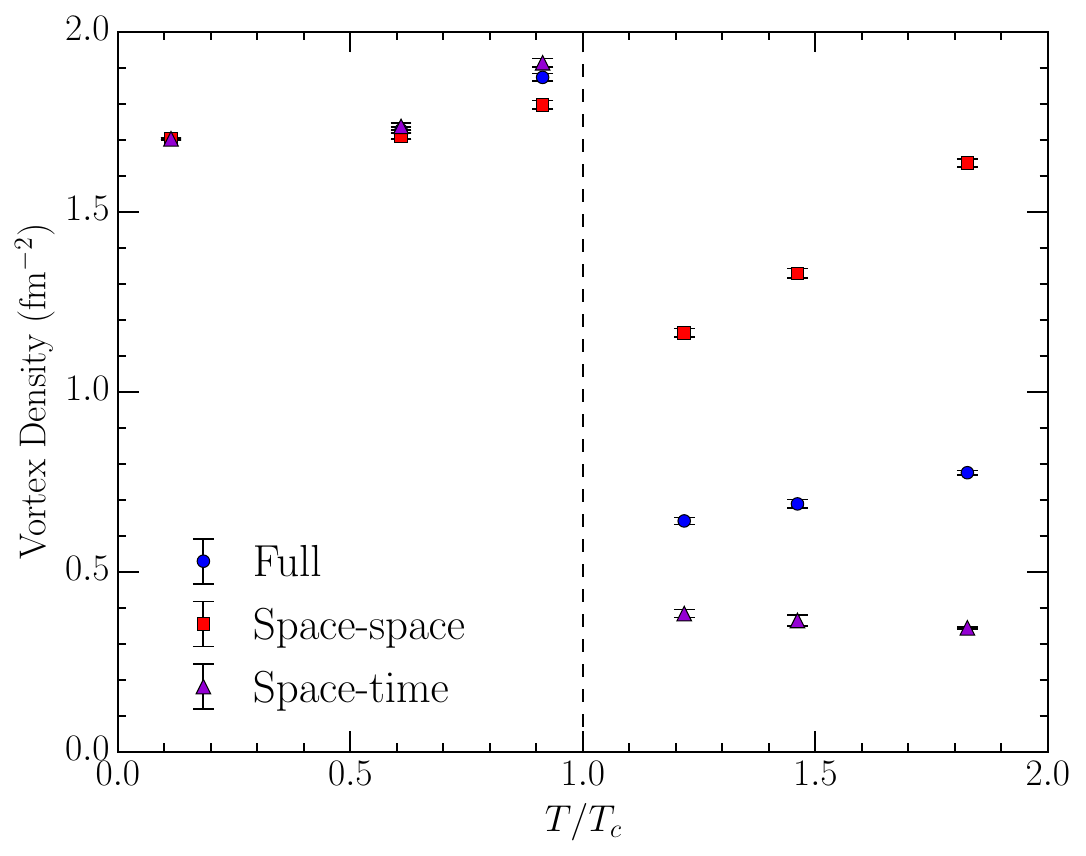}
    \caption{\label{fig:ST_SSdensity} The physical density of space-space and space-time plaquettes pierced in spatial slices of the lattice. There is a greater density in space-time plaquettes approaching the critical temperature from below. Above $T_c$, the space-time density is approximately constant whilst the space-space value increases. Combined, this results in the mild increase in the full density.}
\end{figure}

The evolution of $\rho_\mathrm{vortex}$ with temperature is presented in Fig.~\ref{fig:vortexdensity},
with separate data points for averaging over temporal and spatial slices due to the vastly different structures. As expected, we find a sharp decrease in spatial slices of the lattice as soon as $T_c$ is crossed, coinciding with the absence of a percolating cluster. The area density in the temporal slices also undergoes an initial smaller drop at $T_c$. Both of these subsequently grow in value above $T_c$. This is particularly notable in the temporal slices for which $\rho_\mathrm{vortex}$ proceeds to increase back to below $T_c$ levels. This agrees with the qualitative conclusion reached through the visualisations.

Another curious occurrence is the slight increase in vortex density as $T_c$ is approached from below, as seen with the data points at $T/T_c = 0.913$ in Fig.~\ref{fig:vortexdensity}. These sit clearly above the other points below $T_c$, and is especially pronounced in spatial slices. Greater insight into this behaviour can be obtained by decomposing the vortex density in spatial slices to consider the proportion of space-space and space-time plaquettes pierced separately. The space-space density calculated in this manner is equivalent to the vortex density in temporal slices from Fig.~\ref{fig:vortexdensity}, which comprise entirely space-space plaquettes. After averaging over all slices of the lattice, every space-space plaquette has been accounted for in both quantities. Accordingly, there must be interesting behaviour in the space-time plaquettes that induces the larger density in spatial slices. These decomposed densities are shown in Fig.~\ref{fig:ST_SSdensity}, with the ``full" spatial density from Fig.~\ref{fig:vortexdensity} overlaid for reference.

It is reassuring that the proportion of space-space plaquettes pierced coincides with the temporal density, as required. Fig.~\ref{fig:ST_SSdensity} then explicitly reveals a divergence in the density of space-space and space-time plaquettes pierced approaching the phase transition from below, with a larger density of vortices piercing space-time planes. This asymmetry has previously been seen in SU(2) near $T_c$ \cite{PvortexStructure}, and accounts for the greater vortex density in spatial slices over temporal slices. In addition, the density in space-time plaquettes remains approximately constant above $T_c$, which explains why the increase in the ``full" density in spatial slices is mild in comparison to that found in temporal slices.

It is now clear why the vortex density in spatial slices exceeds that in temporal slices as the phase transition is approached from below. We understand that deconfinement is associated with an alignment of the vortex sheet with the temporal axis, whereas at low temperatures there is no preferred orientation. Thus, before the deconfinement transition can take place, a rearrangement of the vortex sheet is required, and Fig.~\ref{fig:ST_SSdensity} reveals this manifests as a preference to pierce space-time plaquettes as it prepares to align with the temporal axis. These will be features to look for in the branching point statistics to ascertain whether they are recurring characteristics across various vortex attributes.

\subsection{Non-percolating clusters} \label{subsec:secondaryclusters}
Next, we investigate the non-percolating aspects of the centre vortex structure. These encompass any cluster seen in Figs.~\ref{fig:Nt12Vis}--\ref{fig:Nt4Vis} that does not spread through the slice volume, and can be broken down into two distinct categories. First, there are the small structures in temporal slices that contrast the ``primary" percolating cluster at all temperatures; these we specifically refer to as ``secondary" clusters. The second category concerns spatial slices above $T_c$ where percolation is absent; this grouping therefore covers \textit{all} clusters observed in these slices.

\begin{figure}
	\centering
	\includegraphics[width=\linewidth]{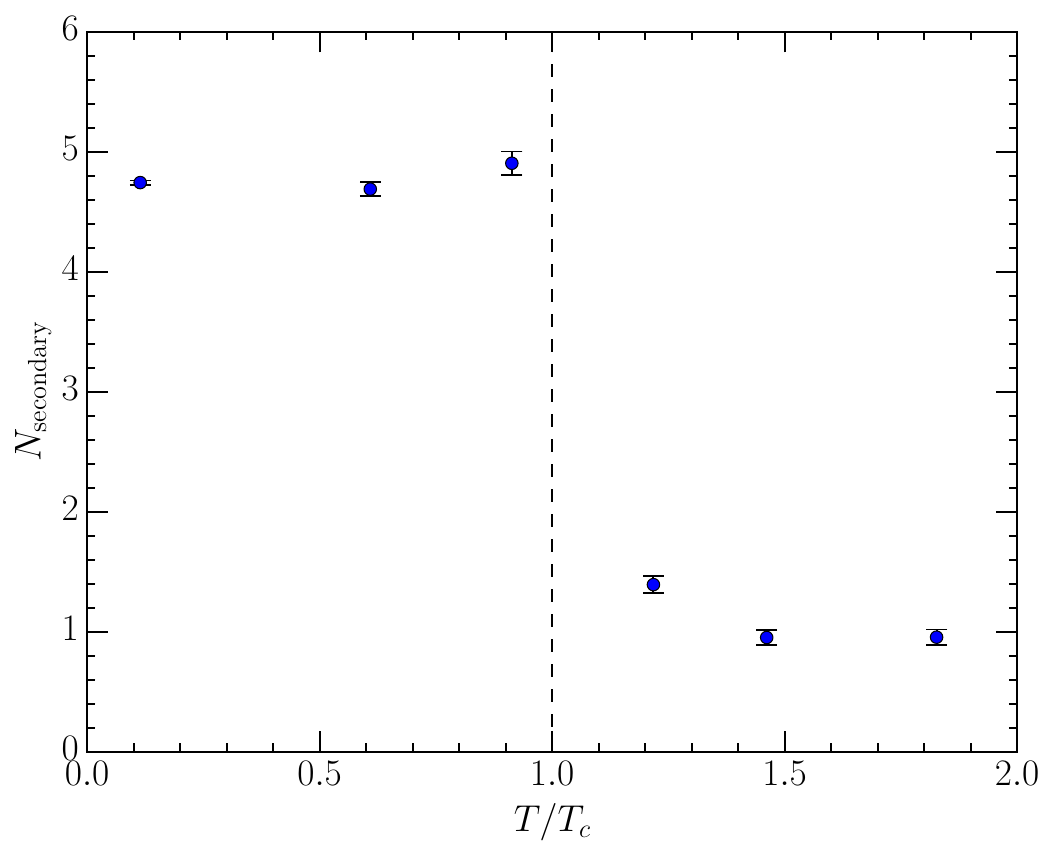}
	\caption{\label{fig:Nsecondary} The temperature-dependence of the average number of secondary vortex clusters $N_\mathrm{secondary}$ in temporal slices of the lattice. There is a marked drop at the critical temperature, after which there is a continuing decrease.}
	
	\vspace{2em}
	
	\includegraphics[width=\linewidth]{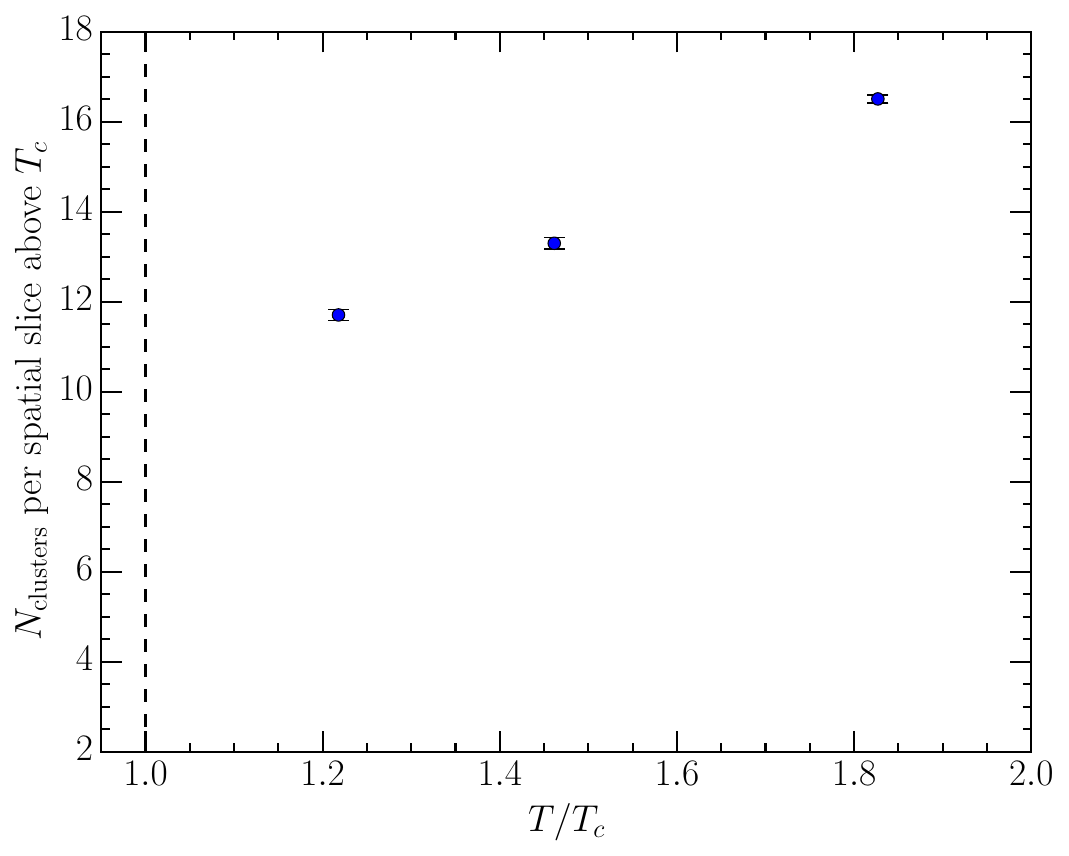}
	\caption{\label{fig:Ndisconnected} The average number of vortex clusters per spatial slice above $T_c$. There is a steady upward trend in the amount of vortex lines as the temperature continues to increase away from $T_c$. This can be directly attributed to the growing vortex density in temporal slices in this regime.}
\end{figure}

To start, the number of secondary clusters $N_\mathrm{secondary}$ is presented for the temporal slices in Fig.~\ref{fig:Nsecondary}.
We note that the notion of ``secondary" clusters also applies to spatial \linebreak slices for $T<T_c$, though our primary focus is directed to temporal slices, which comprise a percolating cluster for all temperatures. This allows an unambiguous exploration on the presence of secondary clusters in the deconfined phase. Figure \ref{fig:Nsecondary} reveals a pronounced drop in the number of secondary clusters as the critical temperature is crossed. This was suspected from the visualisations. It then continues to decline above $T_c$ to an average of only one secondary cluster per slice at our two highest temperatures. The extent to which $N_\mathrm{secondary}$ falls off crossing $T_c$ suggests this is another aspect of vortex geometry that characterises the deconfinement phase transition in the pure-gauge theory. The decrease above $T_c$ may be partly associated with the increase in vortex density illustrated in Fig.~\ref{fig:vortexdensity}. As the volume fills with vortices, there is less space for secondary clusters. That being said, it is currently unclear why high temperatures result in the initial suppression of secondary clusters for $T>T_c$.

Second, we inquire into the number of clusters $N_\mathrm{clusters}$ per spatial slice above $T_c$. This is shown in Fig.~\ref{fig:Ndisconnected}. These are small, mutually disconnected clusters which primarily wind around the temporal dimension, and serve as a basic reflection of the spatial structure seen in time slices. Figure \ref{fig:Ndisconnected} reveals a gradual increase in the number of these clusters with temperature, which was not immediately apparent from the visualisations. This increase is inherently connected to the growing vortex density in temporal slices for $T>T_c$. It follows that when subsequently slicing along a spatial dimension, there are more jets within a given slice. This naturally leads to a greater abundance of vortex lines aligned with the temporal axis in spatial slices.

\section{Branching point geometry} \label{sec:branchingpoints}
Due to the existence of two distinct nontrivial centre phases, SU(3) vortices experience vortex branching where an $m=\pm 1$ vortex splits into two $m=\mp 1$ vortices. This is allowed due to the conservation of centre charge modulo $N$. Branching does not occur in SU(2) where there is only a single nontrivial phase. Therefore, branching points offer a unique avenue of investigation into centre vortices in SU(3) compared to SU(2).

\begin{figure}
	\centering
	\includegraphics[width=0.49\linewidth]{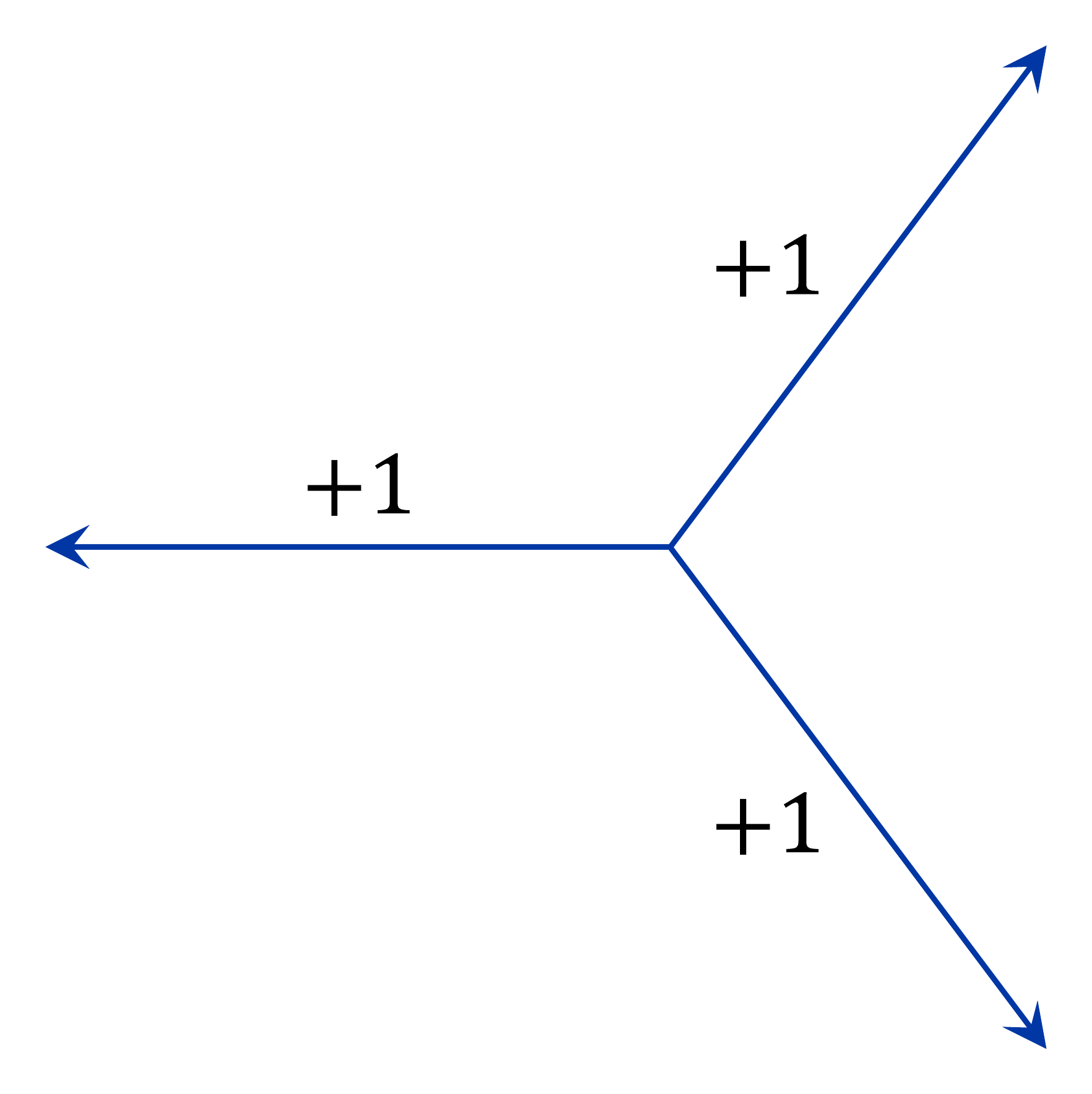}
	\includegraphics[width=0.49\linewidth]{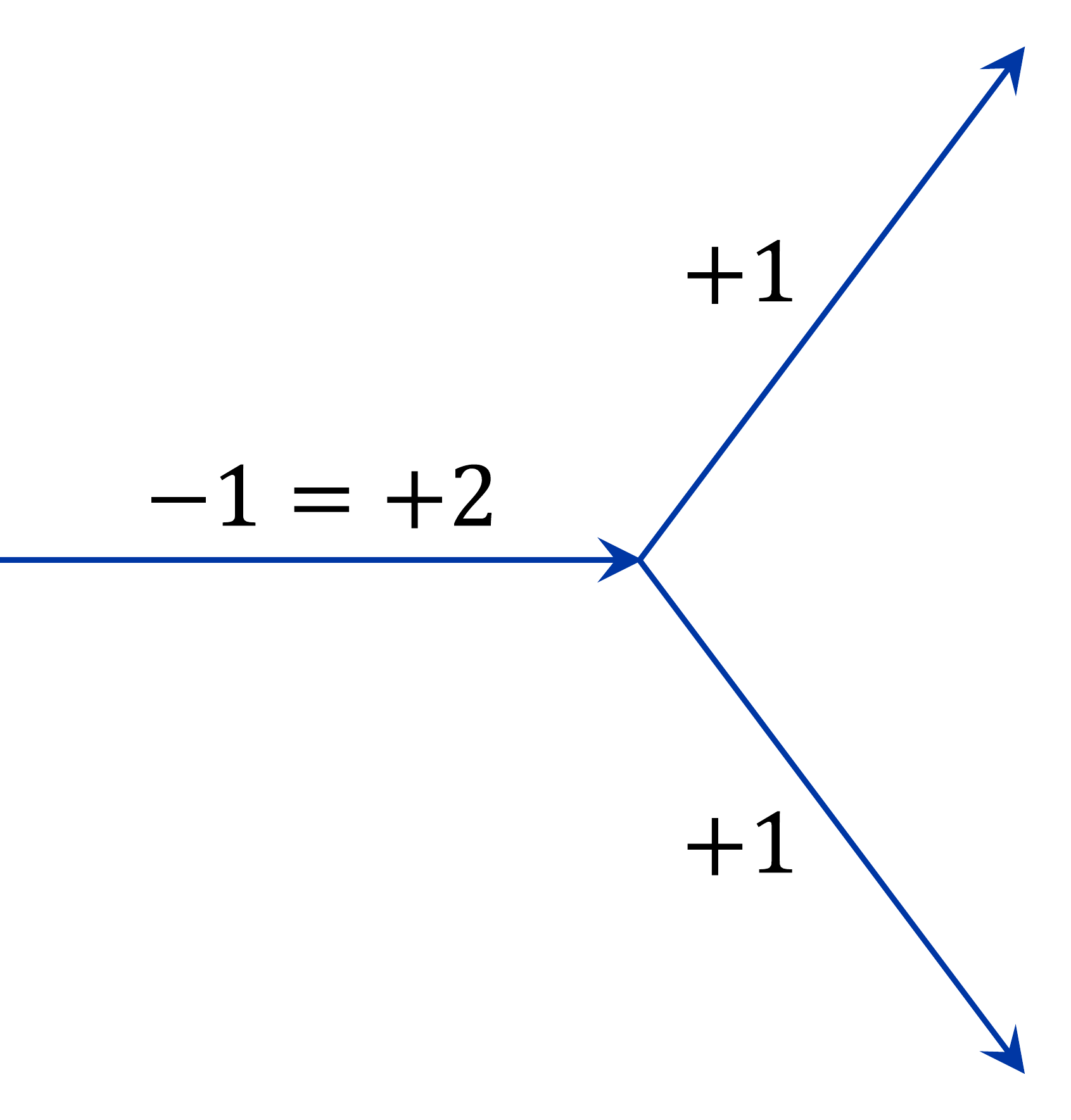}
	\caption{\label{fig:branching} Schematic of a monopole vertex (\textbf{left}) versus a branching point (\textbf{right}). The monopole vertex follows our convention to illustrate the directed flow of $m=+1$ centre charge. Reversal of the left-hand arrow in the diagram shows the flow of charge $m=-1$, as seen on the right. Due to periodicity in the centre charge, $m=-1$ is equivalent to $m=+2$. Thus, the right-hand diagram depicts branching of centre charge.}
\end{figure}

Since reversing the orientation of a jet indicates the flow of the opposite centre charge, branching points can equivalently be interpreted as the monopoles illustrated in Figs.~\ref{fig:Nt12Vis}--\ref{fig:Nt4Vis}. Here, three vortices of the same centre charge emerge from, or converge to, a single point. A schematic of this equivalence between branching and monopole points is given in Fig.~\ref{fig:branching}.
Recalling our visualisations show the flow of $m=+1$ centre charge, branching points exclusively appear as monopoles in our visualisations.

\subsection{Branching point density} \label{subsec:branchingpointdensity}
\begin{figure}
    \centering
    \includegraphics[width=\linewidth]{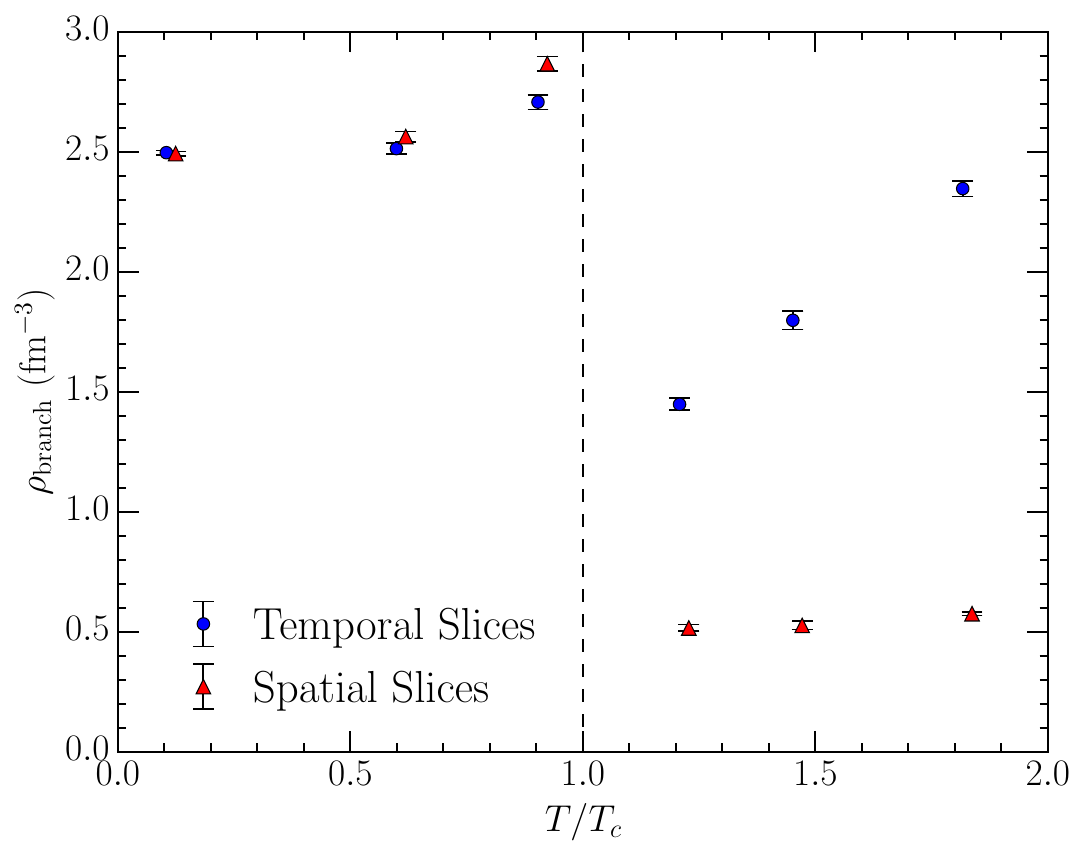}
    \caption{\label{fig:volumebranchingdensity} The density of vortex branching points in temporal and spatial slices of the lattice for each finite-temperature ensemble. The trends in both temporal and spatial slices match those identified for the vortex area density.}
    
    \vspace{2em}
    
    \includegraphics[width=\linewidth]{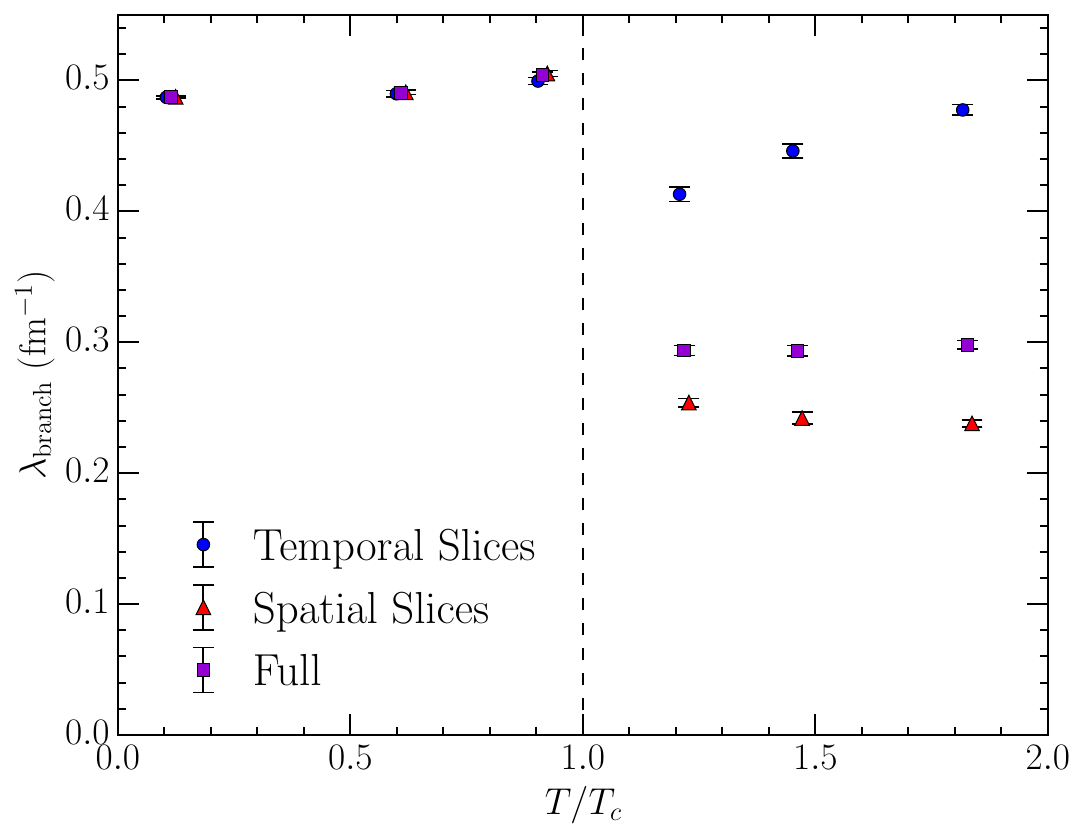}
    \caption{\label{fig:linearbranchingdensity} The linear density of vortex branching points in temporal and spatial slices of the lattice for each finite-temperature ensemble, and a ``full" value from averaging over all four slice dimensions. The trends in temporal and spatial slices coincide with those identified for vortex and branching point densities, though are less pronounced. The full dimensionally averaged value is notably constant above $T_c$.}
\end{figure}

The first quantity of interest is the branching point density, defined for a given three-dimensional slice as the proportion of elementary cubes that contain a branching point,
\begin{equation}
    \hat{\rho}_\mathrm{branch} = \frac{\text{Number of branching points in slice}}{V_\mathrm{slice}} \,.
\end{equation}
As a volume density, the appropriate physical quantity is $\rho_\mathrm{branch} = \hat{\rho}_\mathrm{branch}/a^3$, which is known to scale correctly as $a\to 0$ \cite{ConfinementXI,Branching}. Its evolution with temperature is presented in Fig.~\ref{fig:volumebranchingdensity}.
We find identical patterns to those identified for the vortex area density in both temporal and spatial slices. This includes the peculiar increase near $T_c$, which is again stronger in spatial slices. That being said, the drop in $\rho_\mathrm{branch}$ in spatial slices across $T_c$ is even larger than for $\rho_\mathrm{vortex}$. This is to be expected---with the vortices principally aligned along the short temporal dimension, there are very few opportunities for the vortex line to branch, resulting in a substantial suppression of $\rho_\mathrm{branch}$ in spatial slices.

In temporal slices, the increase in $\rho_\mathrm{branch}$ above $T_c$ is connected to the corresponding behaviour of $\rho_\mathrm{vortex}$, where an increase in vortex density naturally allows for more branching chances within the same spatial volume. In this scenario a more interesting quantity might be the \textit{linear} branching point density,
\begin{equation} \label{eq:linearbranchingdensity}
    \hat{\lambda}_\mathrm{branch} = \frac{\text{Number of branching points in slice}}{\text{Number of vortices in slice}} \,,
\end{equation}
i.e. the proportion of vortices that undergo branching. The corresponding physical quantity is $\lambda_\mathrm{branch} = \hat{\lambda}_\mathrm{branch}/a$. As defined in Eq.~(\ref{eq:linearbranchingdensity}), $\hat{\lambda}_\mathrm{branch}$ can also be interpreted as an estimate of the average probability per unit length for a vortex to branch. We will see in Sec.~\ref{subsec:branchingpointseparations} that this probability is in fact distance dependent.

Following from our definition of $\lambda_\mathrm{branch}$, one finds the below simple relation summarising how various vortex properties intertwine,
\begin{equation} \label{eq:statisticsconnection}
	\rho_\mathrm{branch} = 3\,\rho_\mathrm{vortex}\,\lambda_\mathrm{branch} \,.
\end{equation}
This is derived by a straightforward substitution of Eqs.~(\ref{eq:vortexdensity})--(\ref{eq:linearbranchingdensity}), and noting that the number of vortices in a slice [from the denominator of Eq.~(\ref{eq:linearbranchingdensity})] is by definition equal to the number of nontrivial plaquettes [from the numerator of Eq.~(\ref{eq:vortexdensity})]. Thus, a constant $\lambda_\mathrm{branch}$ would give a clear-cut relationship between the vortex and volume branching point densities. In contrast, if $\lambda_\mathrm{branch}$ is also subject to nontrivial evolution with temperature the dependency becomes more complicated. We plot the linear branching point density in temporal and spatial slices as a function of temperature in Fig.~\ref{fig:linearbranchingdensity}. This is overlaid by a ``full" density $\bar{\lambda}_\mathrm{branch}$ obtained by averaging $\lambda_\mathrm{branch}$ over all four slice dimensions,
\begin{equation}
	\bar{\lambda}_\mathrm{branch} = \frac{1}{4} \sum_\mu \lambda_{\mathrm{branch}}(\mu) \,,
\end{equation}
where $\lambda_{\mathrm{branch}}(\mu)$ is the linear branching density for slice dimension $\mu$. This is relevant to the following discussion.

The same general trends are found in $\lambda_\mathrm{branch}$ as for $\rho_\mathrm{vortex}$ and $\rho_\mathrm{branch}$ but visibly subdued, with a smaller drop at $T_c$ for both temporal and spatial slices. Importantly, the fact $\lambda_\mathrm{branch}$ increases in temporal slices above $T_c$ implies the corresponding increase in $\rho_\mathrm{branch}$ is not solely due to the growing vortex density. It is additionally caused in part by an increase in the inherent fraction of vortices that undergo branching, as per Eq.~(\ref{eq:statisticsconnection}).

We also find that the ``full" dimensionally averaged value $\bar{\lambda}_\mathrm{branch}$, forgoing the small increase near $T_c$, remains approximately constant either side of the phase transition (over our temperature range). This value changes when crossing $T_c$, with the reduced value in the high-temper\-ature phase signifying a corresponding change in the vortex geometry. This can be understood by noting that the shift in vortex geometry to principally align with the temporal axis implies there is generally one less dimension available for a vortex to branch into, explaining why $\bar{\lambda}_\mathrm{branch}$ decreases through the phase transition. For a simple heuristic argument, the collapse in vortex geometry \linebreak from four- to three-dimensional implies the number of available dimensions for a vortex line within the sheet to branch into decreases from three to two. It follows that at leading order, one might expect the value of $\bar{\lambda}_\mathrm{branch}$ to drop by a factor of $2/3$ at $T_c$. Figure \ref{fig:linearbranchingdensity} shows reasonable consistency with this proposal, for which taking a ratio between the average values in the two phases yields $\approx 0.60$.

The fact $\bar{\lambda}_\mathrm{branch}$ subsequently remains constant above $T_c$, despite the significant differences in spatial and temporal structures, provides a baseline that underlies many of our prior findings at high temperature. The vortex correlation measure (Fig.~\ref{fig:correlation}) indicated that the alignment of the vortex sheet with the temporal dimension becomes stronger as the temperature increases above $T_c$. With $\bar{\lambda}_\mathrm{branch}$ constant, this implies that proportionally more of the branching points must occur in temporal slices over spatial slices.

To understand this, recall the alignment manifests in spatial slices as short vortex lines orthogonal to space-space plaquettes, and therefore parallel to the temporal dimension. The occurrence of a branching point would necessitate the piercing of space-time plaquettes, which is suppressed as the alignment becomes stronger. Hence, above $T_c$ the linear branching density decreases in spatial slices, counterbalanced by an increase in temporal slices such that the full density remains constant (as seen in Fig.~\ref{fig:linearbranchingdensity}). This evolution naturally effects an increase in the vortex density in temporal slices, where more plaquettes are pierced from the greater proportion of branching vertices. From Eq.~(\ref{eq:statisticsconnection}), there is consequently a compounding effect that results in a substantial increase in the volume density of branching points.

In summary, the constant value of $\bar{\lambda}_\mathrm{branch}$ above $T_c$ provides an underlying explanation for why the vortex and branching point densities grow in temporal slices of the lattice in the deconfined phase, in spite of the initial drop they experience across the phase transition. Furthermore, the fact that $\bar{\lambda}_\mathrm{branch}$ remains constant whilst the asymmetry between the temporal and spatial extents of the lattice increases with temperature highlights the significance of the vortex geometry in the context of all four dimensions. The geometric nature of the phase transition is understood as a reduction in the dimensionality of vortex percolation from the full four-dimensional lattice to within a three-dimensional submanifold. The primary effect of increasing the temperature in the deconfined phase is an evaporation of branching points from spatial slices as the vortex geometry aligns with the temporal axis, with a commensurate condensation of branching points on temporal slices in exchange such that the average linear branching probability remains constant.

\subsection{Branching point separations} \label{subsec:branchingpointseparations}
In Sec.~\ref{subsec:branchingpointdensity} the bulk properties of branching points were investigated, which we presently extend to a detailed analysis of the intrinsic distribution of branching points throughout the vortex structure. A model for vortex branching proposes that a vortex line has a fixed probability of branching as it propagates through spacetime \cite{Branching}. It follows that the path lengths between consecutive branching points on the lattice would be described by the geometric distribution,
\begin{equation} \label{eq:geodist}
	\Pr\,(k) = p\,(1-p)^{k-1} \,,
\end{equation}
for branching probability $p$ and number of trials $k$.

This has previously been tested at zero temperature in Ref.~\cite{StructureDynamical} by devising an algorithm to determine the separations between branching points and producing a histogram of the results. The conjecture was supported for separations $>3$ where it was found the distribution is approximately exponential. This is spoiled at short distances due to a tendency for branching points to cluster near each other. It is for this reason $\hat{\lambda}_\mathrm{branch}$ defined in Eq.~(\ref{eq:linearbranchingdensity}) fails to provide a reliable estimate of the branching probability.

To determine the probability $p$, a linear function
\begin{equation} \label{eq:linear}
	f(k) = \alpha - \beta k
\end{equation}
is fit to the log of the distribution for separations $k>3$, which by comparing to the logarithm of Eq.~(\ref{eq:geodist}) allows one to extract the corresponding long-range branching probability from the slope parameter $\beta$ as
\begin{equation} \label{eq:problinear}
	p = 1 - e^{-\beta} \,.
\end{equation}
We utilise the algorithm of Ref.~\cite{StructureDynamical} to explore how, if at all, this geometry of branching points and the associated branching probability depends on temperature.

\begin{figure*}
	\centering
	\includegraphics[width=0.48\linewidth]{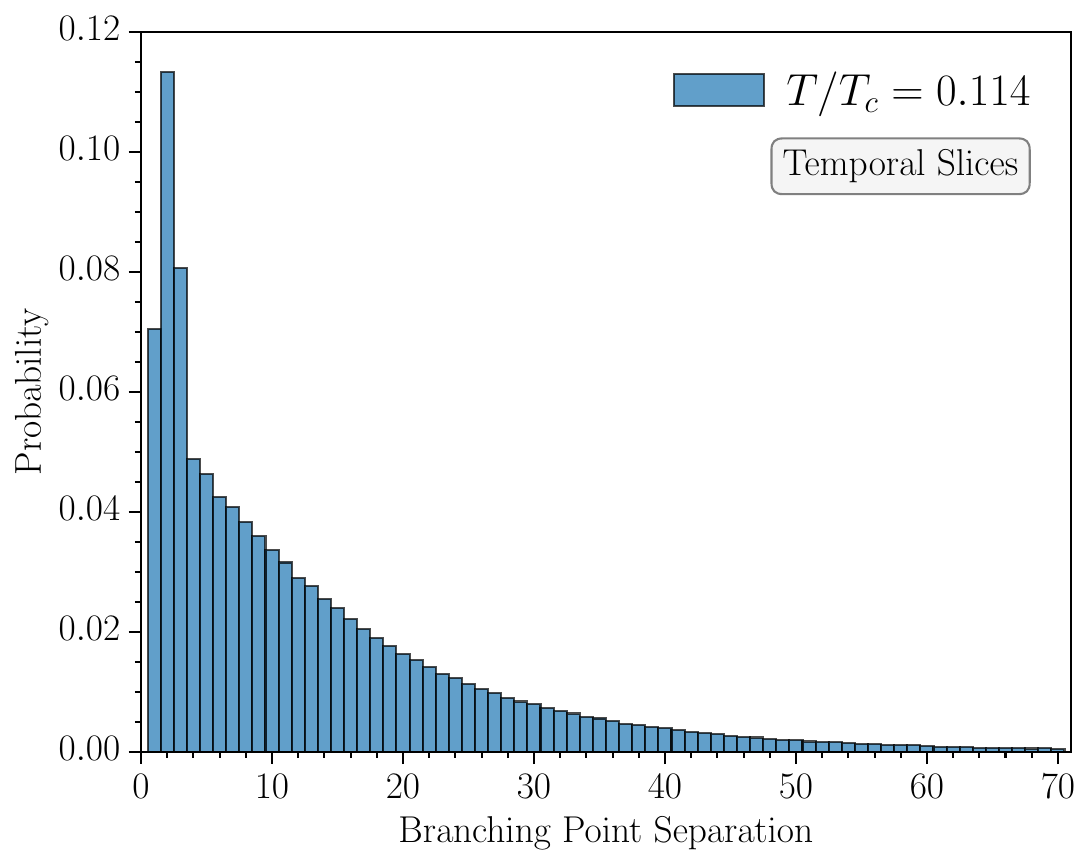}
	\includegraphics[width=0.48\linewidth]{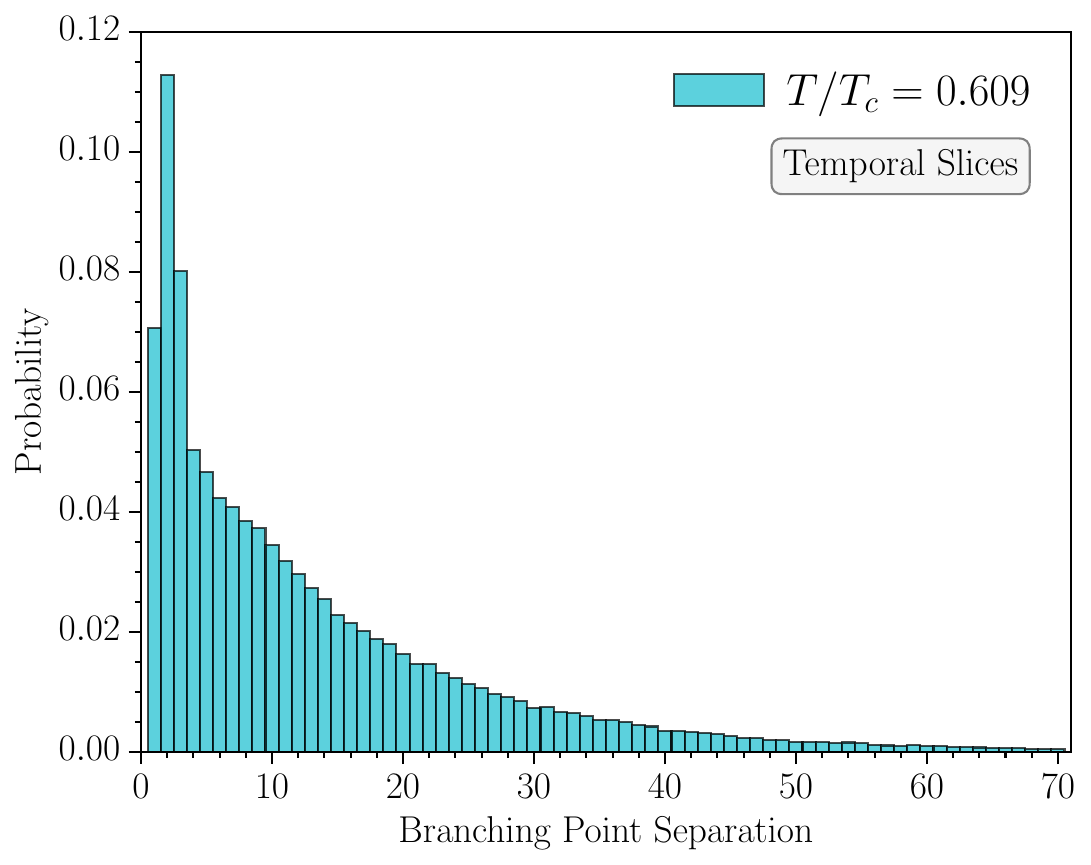}
	
	\vspace{1em}
	
	\includegraphics[width=0.48\linewidth]{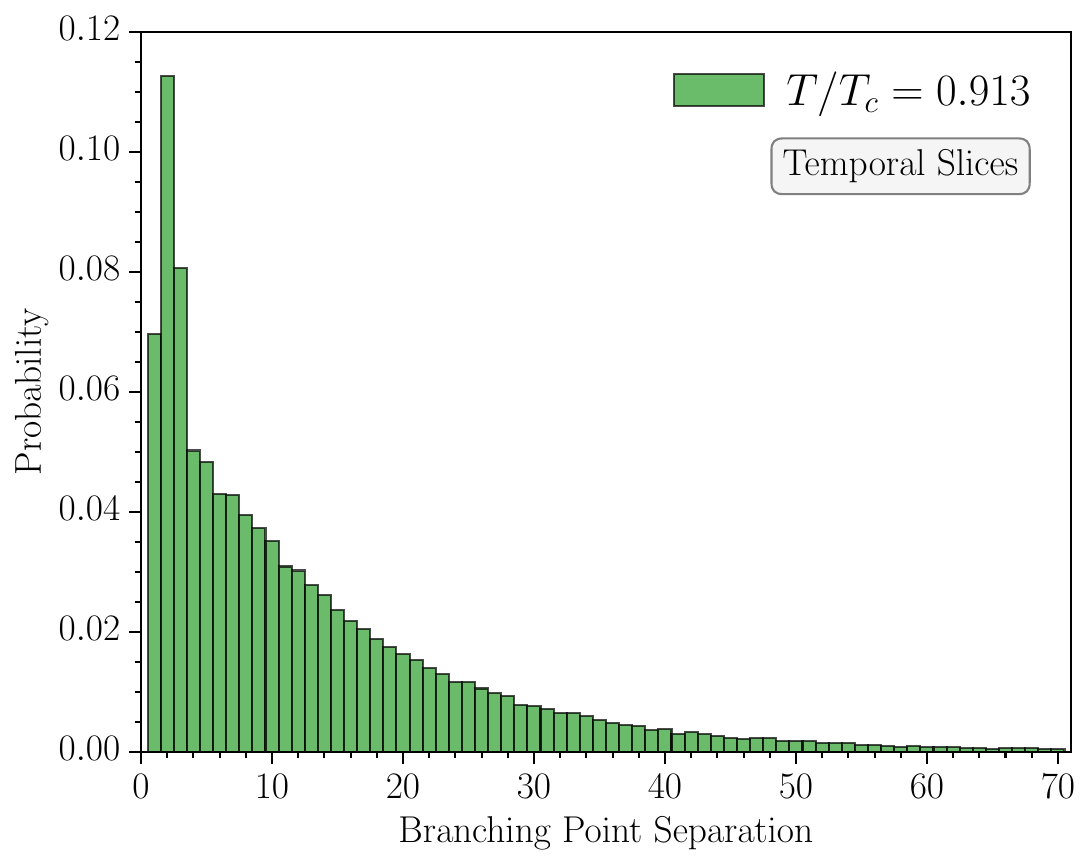}
	\includegraphics[width=0.48\linewidth]{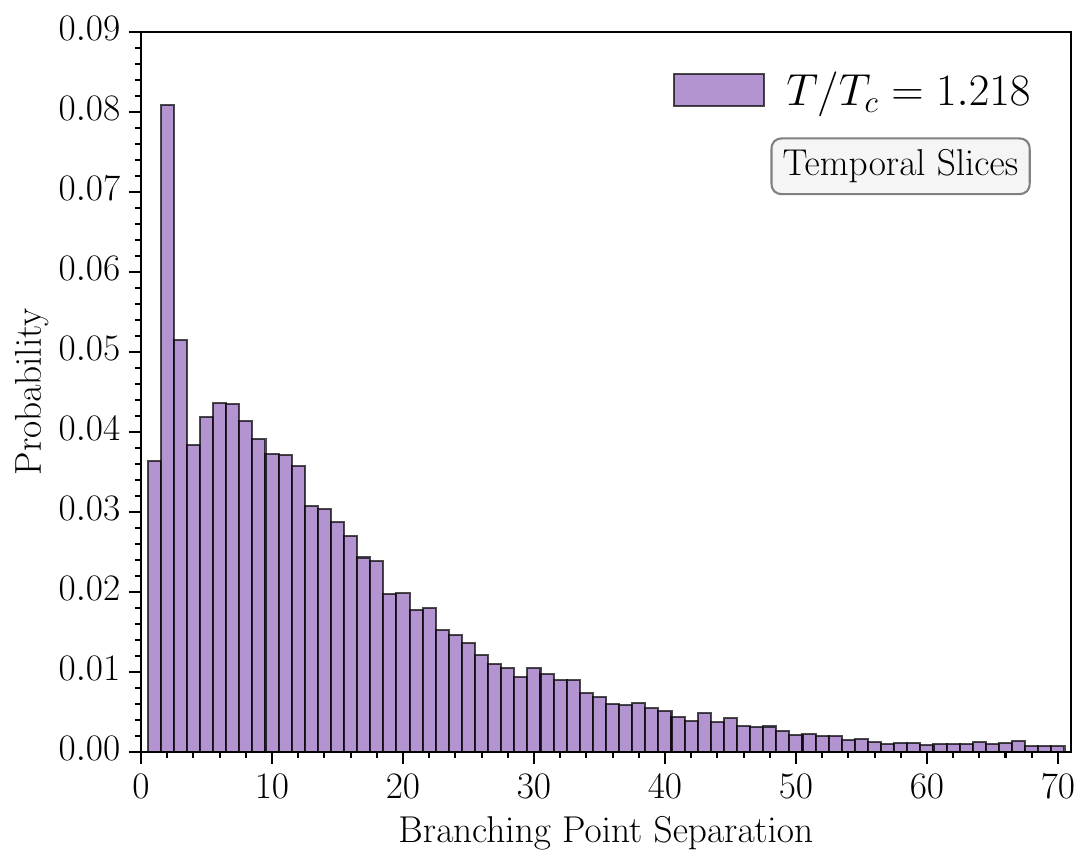}
	
	\vspace{1em}
	
	\includegraphics[width=0.48\linewidth]{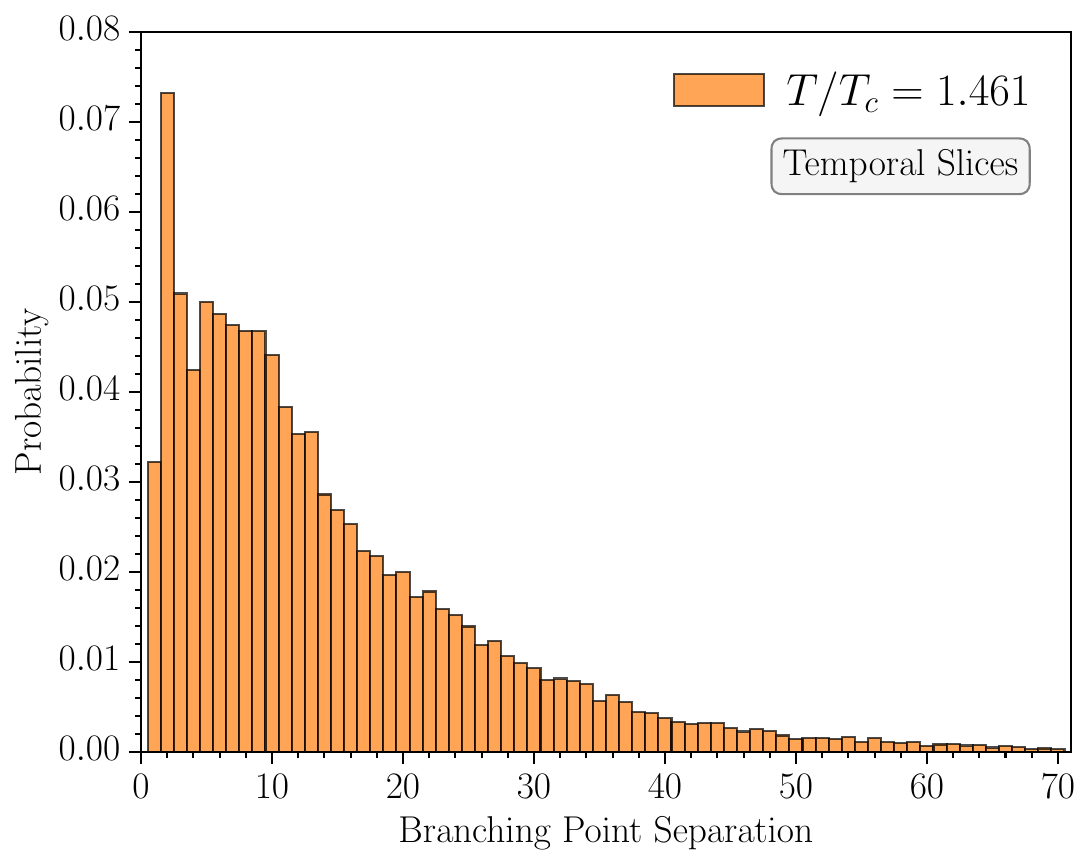}
	\includegraphics[width=0.48\linewidth]{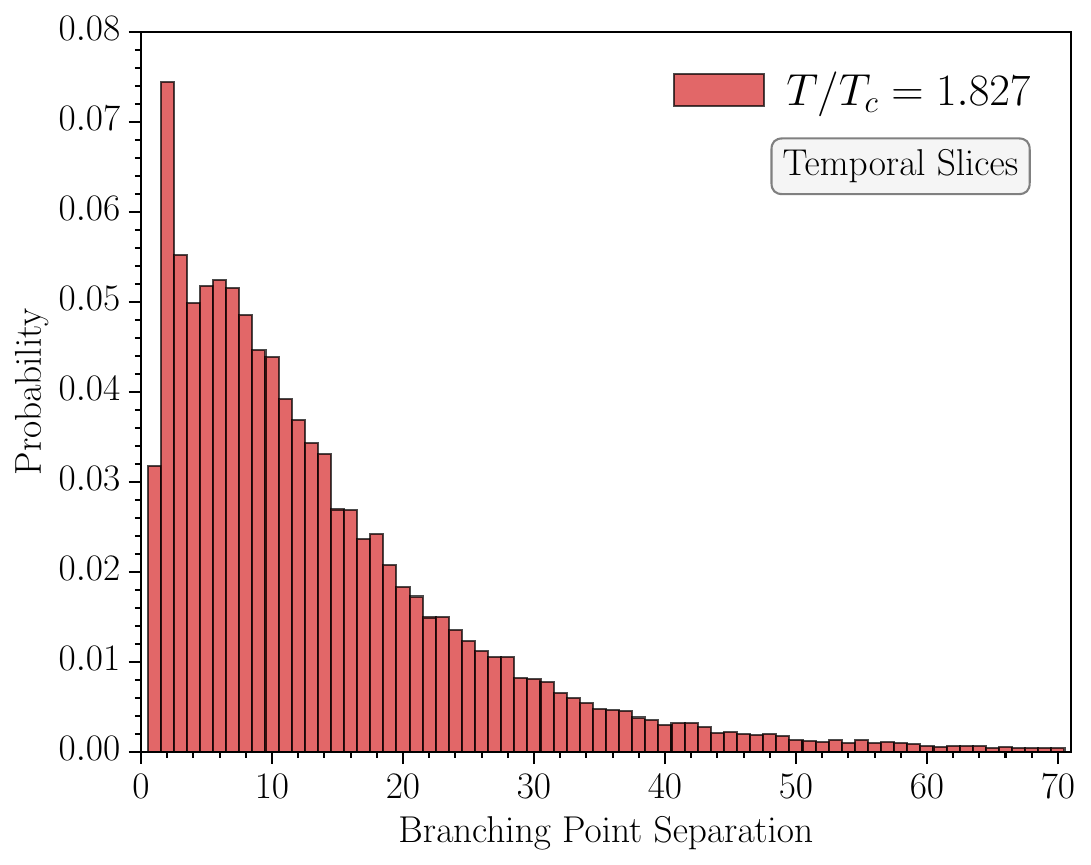}
	\caption{\label{fig:temporalbpsep} The distribution of branching point separations, $k$, in temporal slices of the lattice for each finite-temperature ensemble. The histograms are normalised to unit probability. Below $T_c$ the distributions are similar, exhibiting a clear clustering at separations $k \leq 3$ with a smooth exponential falloff. Above $T_c$ the clustering is visibly softened, and there is also a reduction in probability at separations $k=4$ and $5$.}
\end{figure*}

\begin{figure*}
	\centering
	\includegraphics[width=0.48\linewidth]{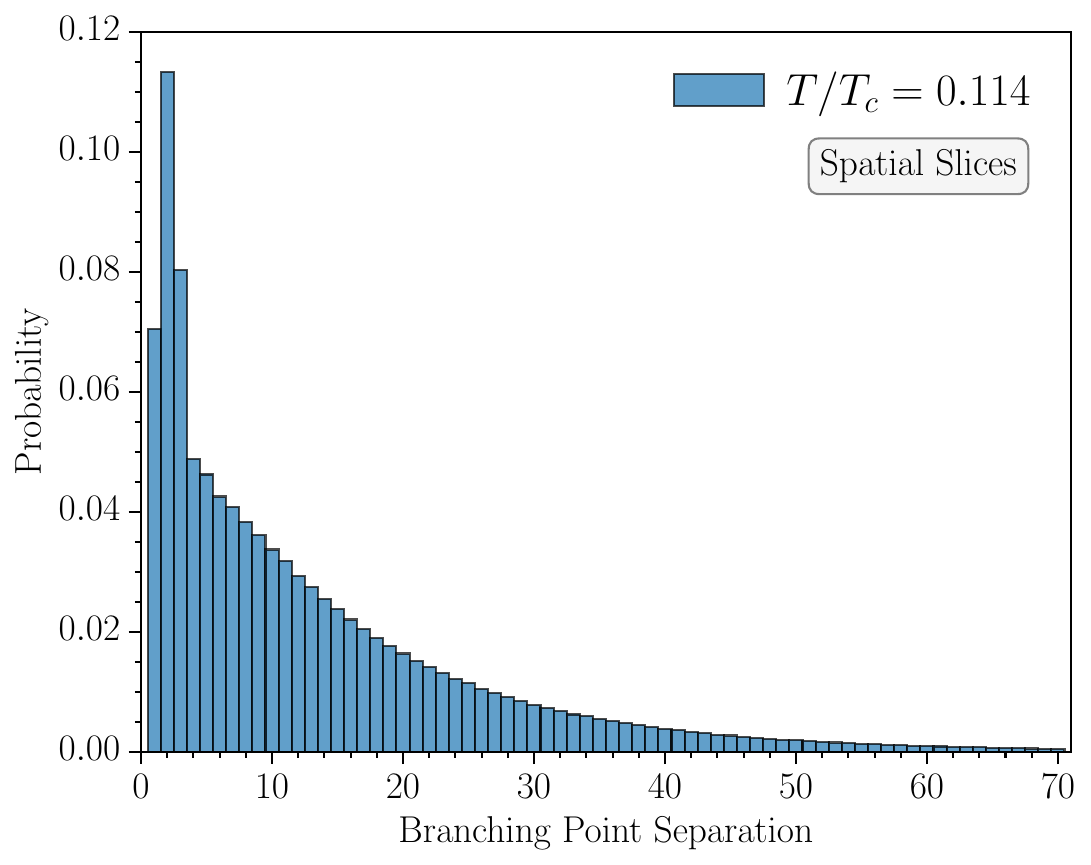}
	\includegraphics[width=0.48\linewidth]{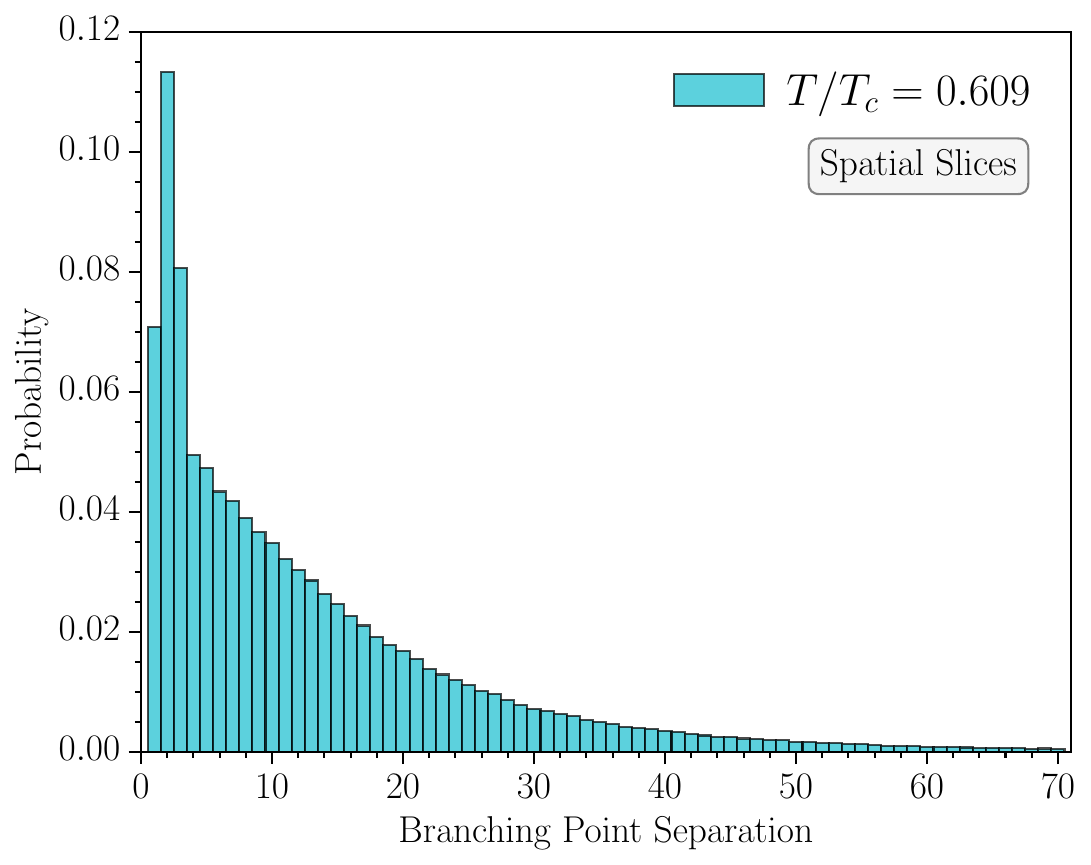}
	
	\vspace{1em}
	
	\includegraphics[width=0.48\linewidth]{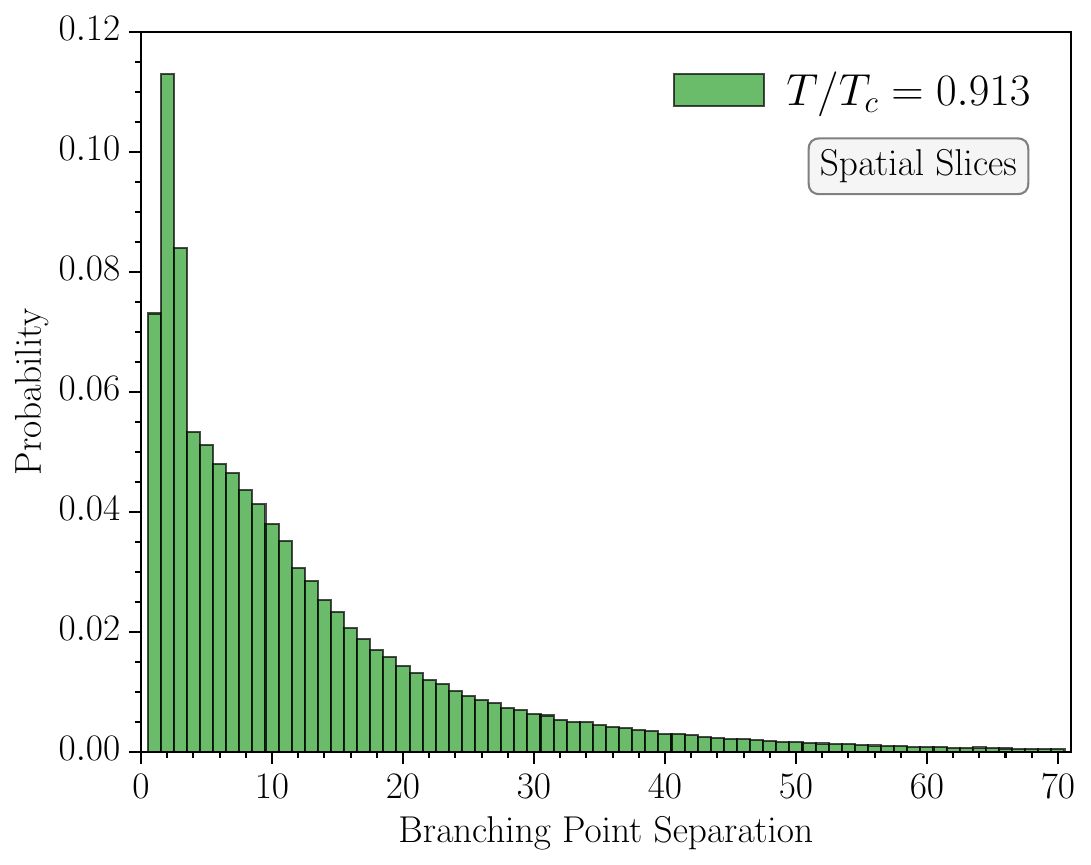}
	\includegraphics[width=0.48\linewidth]{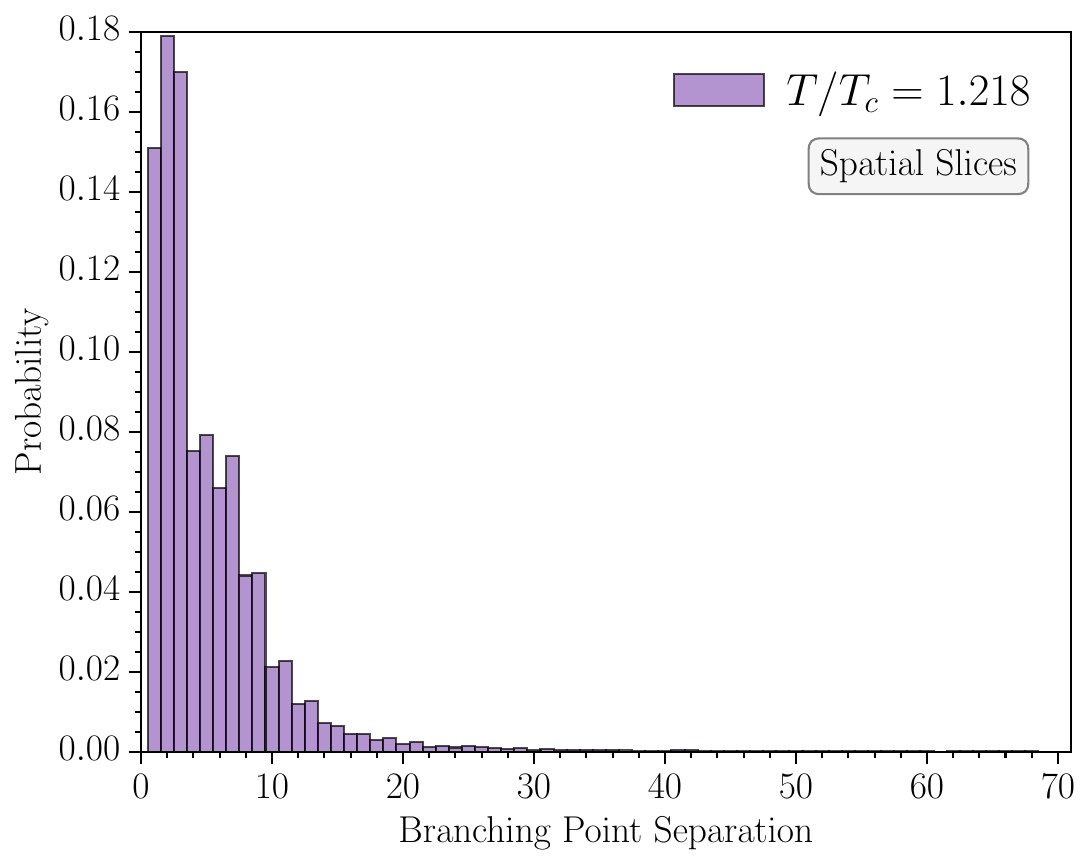}
	
	\vspace{1em}
	
	\includegraphics[width=0.48\linewidth]{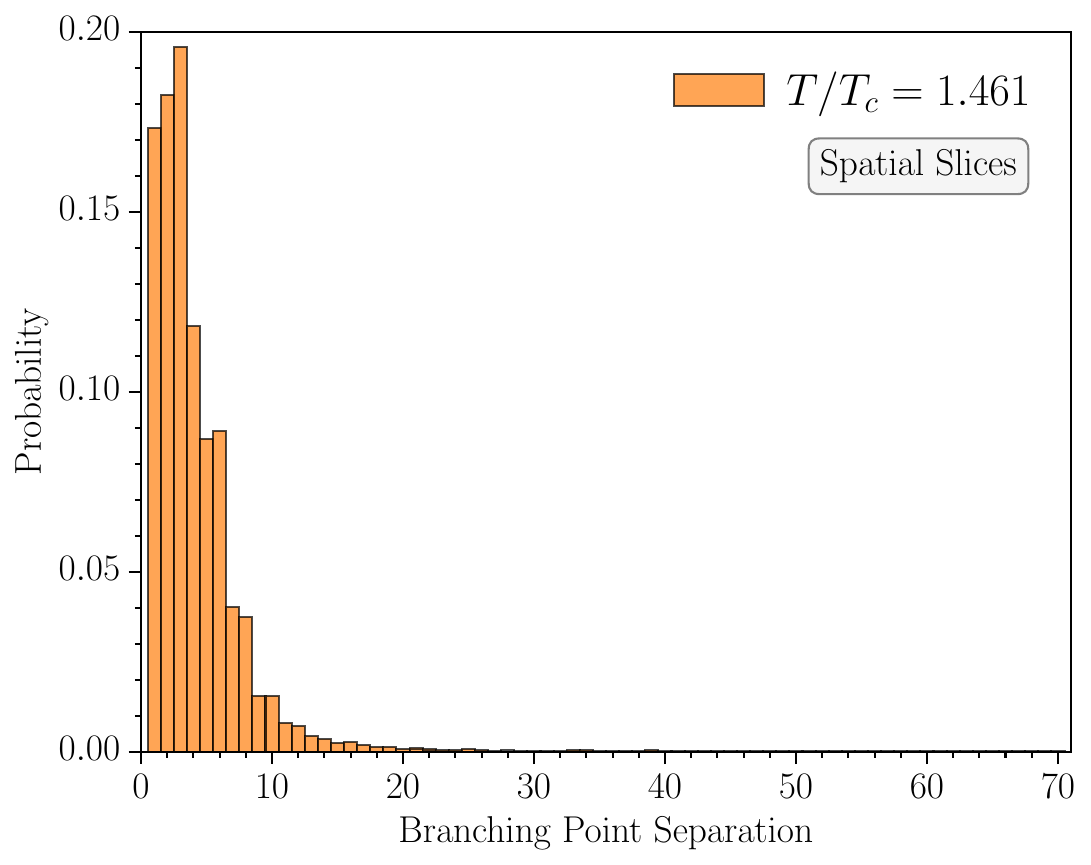}
	\includegraphics[width=0.48\linewidth]{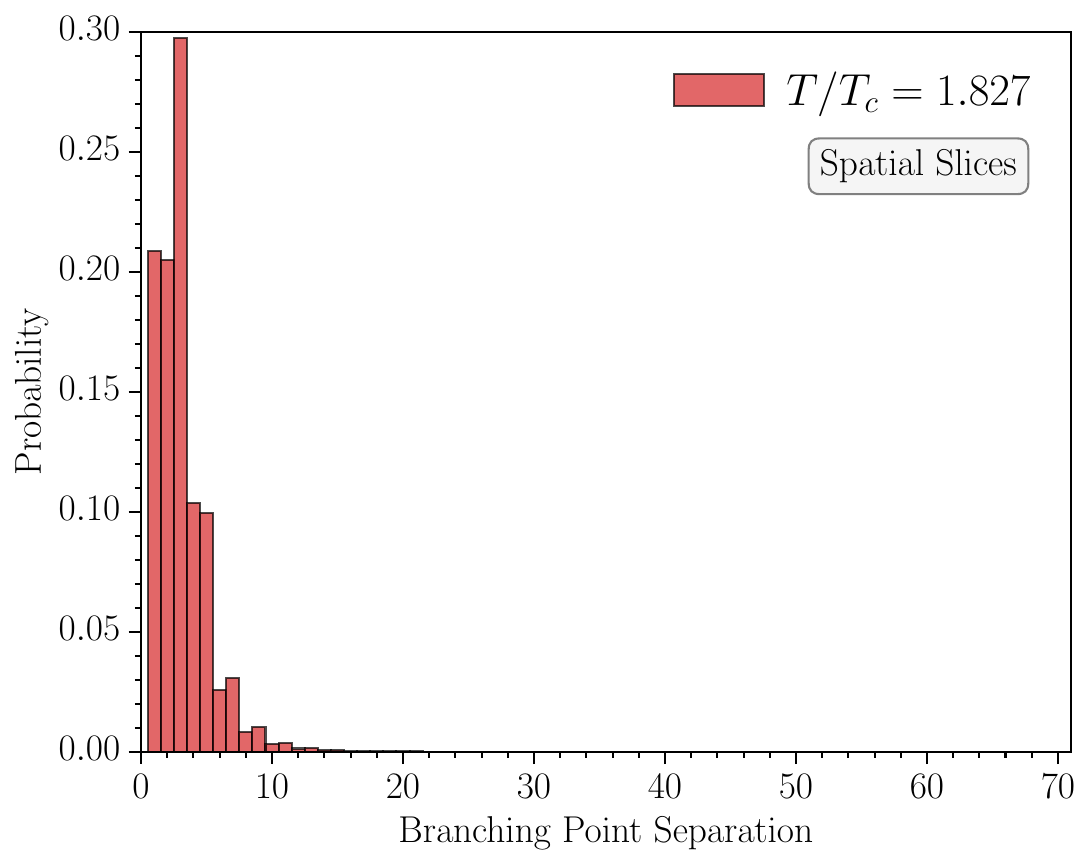}
	\caption{\label{fig:spatialbpsep} The distribution of branching point separations, $k$, in spatial slices of the lattice for each finite-temperature ensemble. The histograms are normalised to unit probability. Below $T_c$, the distributions coincide with that seen in the temporal slices. Above $T_c$, the probability of short separations is increased owing to the alignment of vortices along the temporal dimension, meaning separations $k>N_t$ become increasingly rare.}
\end{figure*}

These distributions are shown independently for temporal slices in Fig.~\ref{fig:temporalbpsep}
and spatial slices in Fig.~\ref{fig:spatialbpsep}.
In the confined phase, the distributions for both temporal and spatial slices are all extremely similar, each featuring the aforementioned clustering at branching point separations $\leq 3$ before a steep falloff in probability leads into a smooth long-range exponential trend.

Above $T_c$, the branching point geometry for temporal and spatial slices diverges, with opposing trends to be found. In spatial slices, the probability of small separations only becomes more pronounced, as is apparent by the height of the respective histogram bins in Fig.~\ref{fig:spatialbpsep}. However, this is presumably an artificial change that arises from the absence of percolation in the deconfined phase, and the alignment of the vortex sheet with the temporal dimension. As seen in the visualisations (Figs.~\ref{fig:Nt6Vis}--\ref{fig:Nt4Vis}), the spatial slices therefore consist of many short vortex lines parallel to the temporal axis.

Consequently, large branching point separations are suppressed, with path lengths $\gtrsim N_t$ becoming increasingly rare. Owing to the very low statistics at large separations, and the sporadic distribution at short distances, performing the aforementioned fits to our spatial slices data above $T_c$ is infeasible. As a result, we refrain from estimating the branching probability in spatial slices. We propose that a possible method to overcome this limitation is to introduce anisotropy into the lattice along the temporal dimension. At high temperatures, this would allow for more branching chances in spatial slices and permit greater distances (in lattice units) between successive branching points. The fits might then successfully be performed over an adequate range of separations.

In contrast, the clustering in temporal slices is diminished in the deconfined phase and continues to subside as the temperature grows. Moreover, this decline in probability at small distances extends to separations $k=4$ and $5$, where above $T_c$ we have $\Pr\,(4) < \Pr\,(5) < \Pr\,(6)$. The familiar exponential decay then takes effect for $k\geq 6$, which is evident from Fig.~\ref{fig:temporalbpsep}. This time the lack of a constant branching probability at short distances is no longer due to a \textit{clustering}, but instead due to a \textit{dilution} of branching points at short- to mid-range scales. Given the temporal centre vortex structure comprises a percolating cluster in both phases, we can be confident this is a genuine shift in the intrinsic arrangement of branching points, rather than arising from an external change in vortex geometry. For instance, the branching point densities in temporal slices are extremely similar at our highest temperature $T/T_c=1.827$ compared to below $T_c$ (Figs.~\ref{fig:volumebranchingdensity} and \ref{fig:linearbranchingdensity}), though the inherent distribution of these branching points is plainly very different.

In Ref.~\cite{StructureDynamical}, it was questioned whether the clustering radius at low temperatures is a physical effect, or a discretisation artefact of a nonzero lattice spacing. Based on the above discussion, we propose it is a physical effect that vanishes at high temperatures (in temporal slices).

Having carefully analysed these distributions, we now proceed to determine the branching probability as a function of temperature. To avoid uncertainty asymmetry induced by taking the log of the data, we utilise a slightly different method to Ref.~\cite{StructureDynamical} and instead perform a direct exponential fit,
\begin{equation} \label{eq:exp}
	f(k) = \zeta \,e^{-\lambda k} \,,
\end{equation}
to the \textit{raw} counts, where $\zeta$ is a normalisation parameter. The connection between the exponential and geometric distributions (see Appendix~\ref{app:expgeo}) allows us to estimate the branching probability for our discrete data as
\begin{equation} \label{eq:probexp}
	p_\mathrm{branch} = 1 - e^{-\lambda} \,,
\end{equation}
cf. Eq.~(\ref{eq:problinear}). To be precise in our methods, we initially perform the linear fit to the logarithm of the counts. The output of this fit is then used to provide accurate initial guesses to the nonlinear exponential fit, $\zeta=e^\alpha$ and $\lambda=\beta$. We find the exponential fit converges to a marginally (but visibly) different optimal solution. This is taken as our best estimate of the model.

\begin{figure*}
	\centering
	\includegraphics[width=0.48\linewidth]{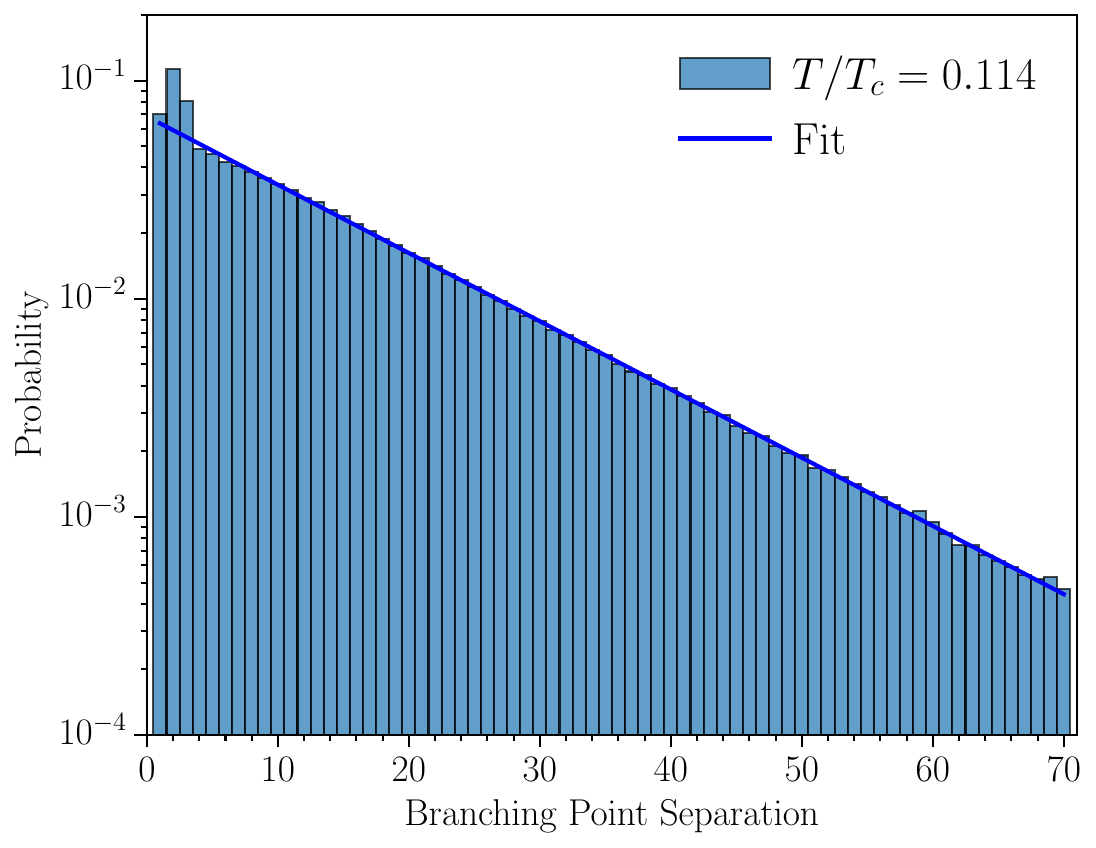}
	\includegraphics[width=0.48\linewidth]{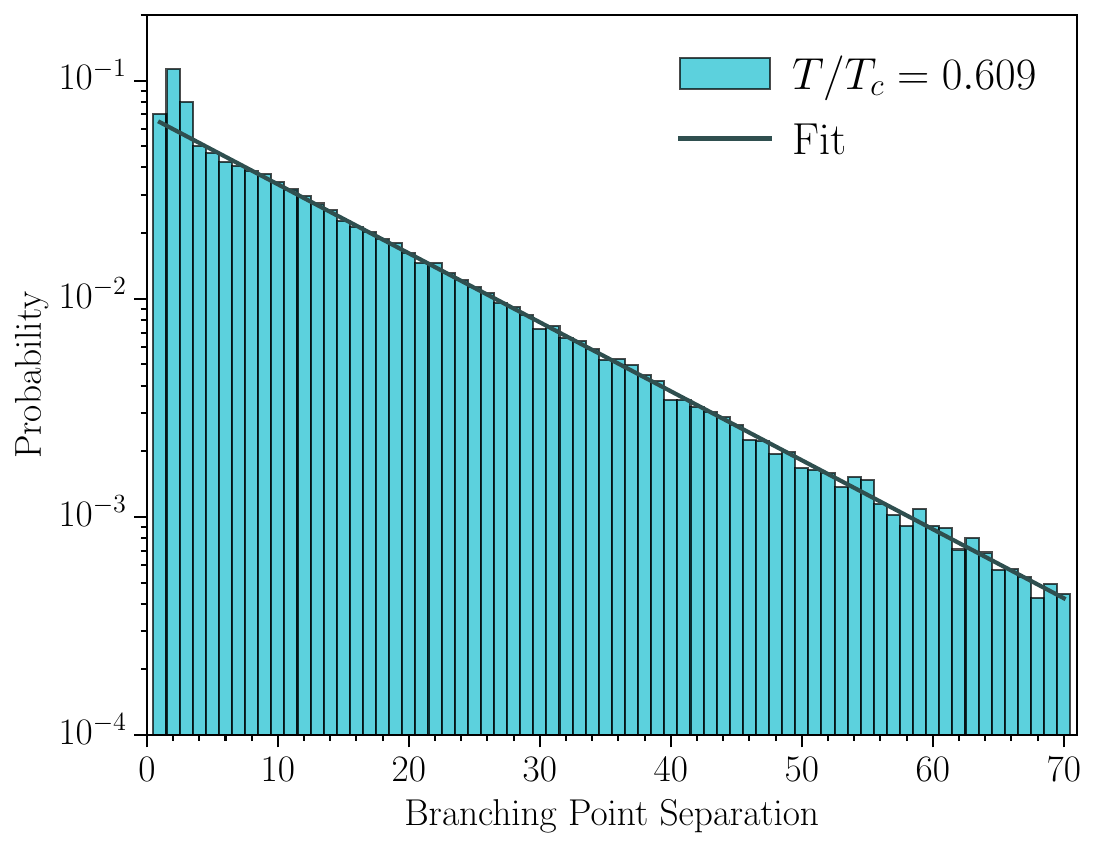}
	
	\vspace{1em}
	
	\includegraphics[width=0.48\linewidth]{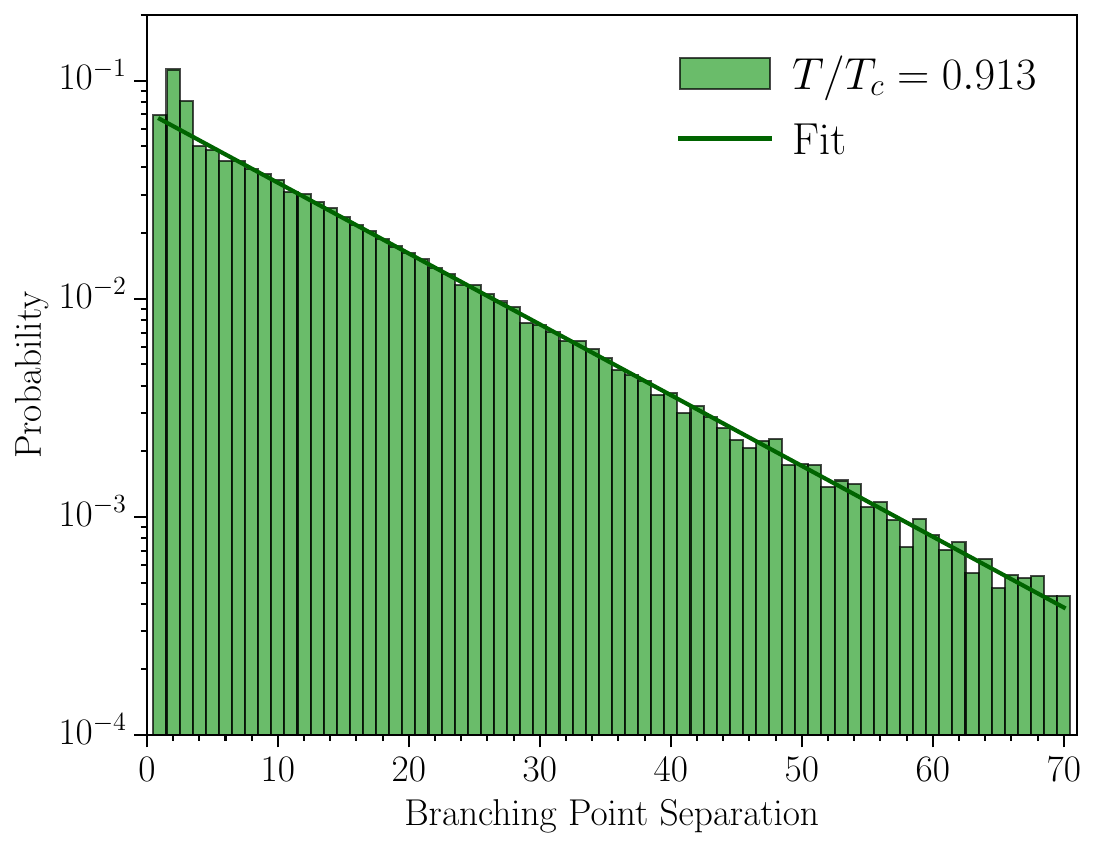}
	\includegraphics[width=0.48\linewidth]{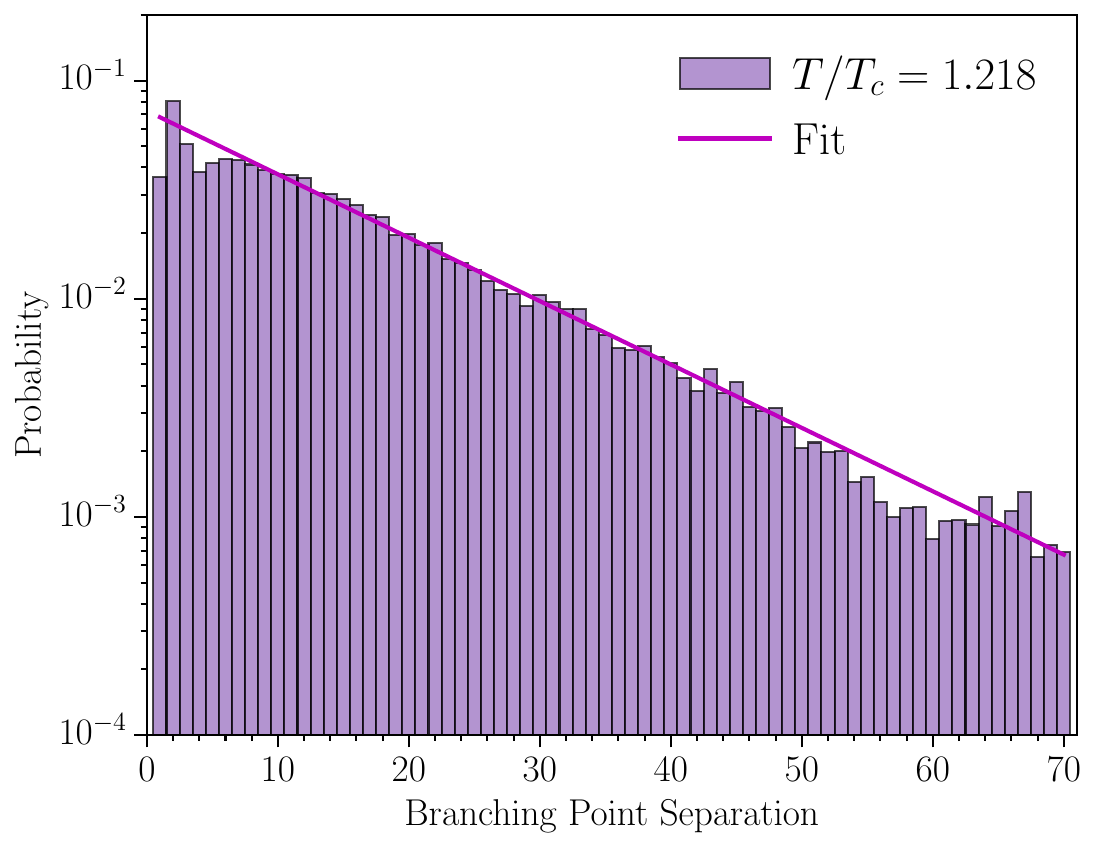}
	
	\vspace{1em}
	
	\includegraphics[width=0.48\linewidth]{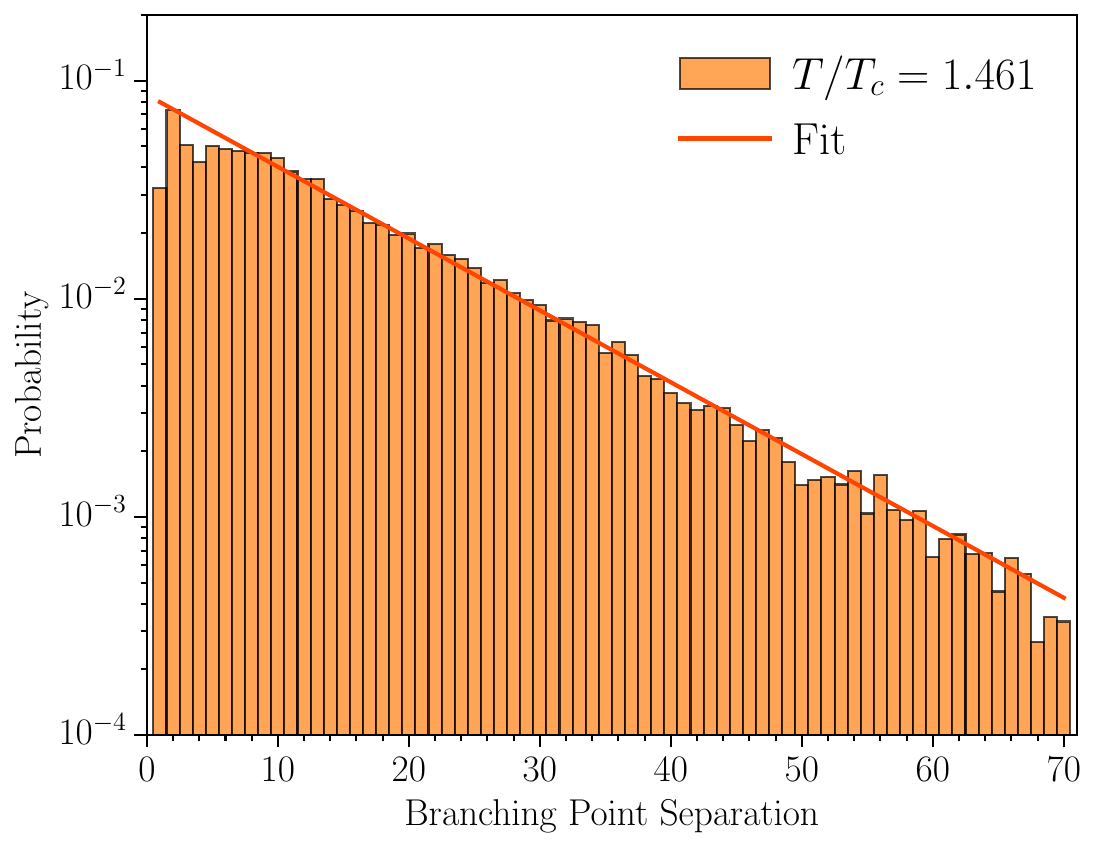}
	\includegraphics[width=0.48\linewidth]{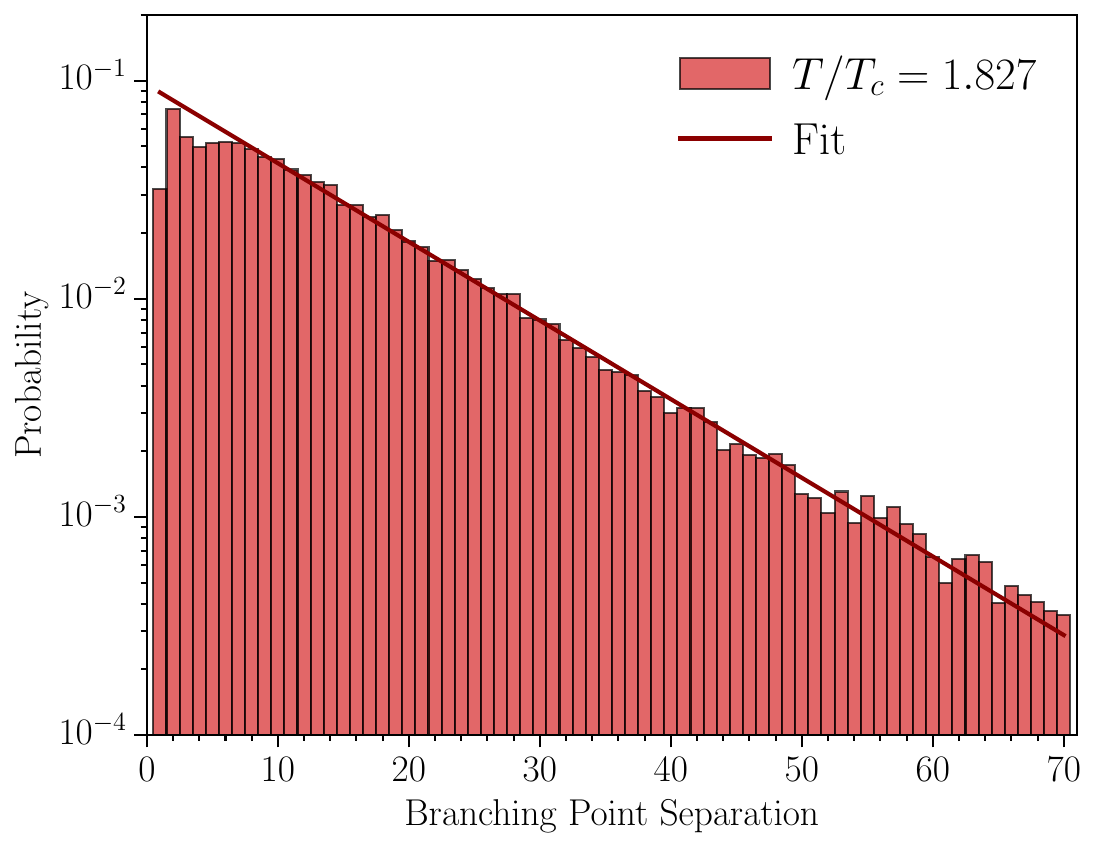}
	\caption{\label{fig:bpsepfits} The distribution of branching point separations, $k$, on a logarithmic scale in temporal slices of the lattice for each finite-temperature ensemble. The exponential fits are overlaid on the histograms, and evidently match the data well barring at very high separations where there is insufficient statistics. The reduction in density at short distances above $T_c$ is visually clear in these plots, with most of the probabilities for separations $k\leq 5$ sitting below the fits.}
\end{figure*}

\begin{figure}
	\centering
	\includegraphics[width=\linewidth]{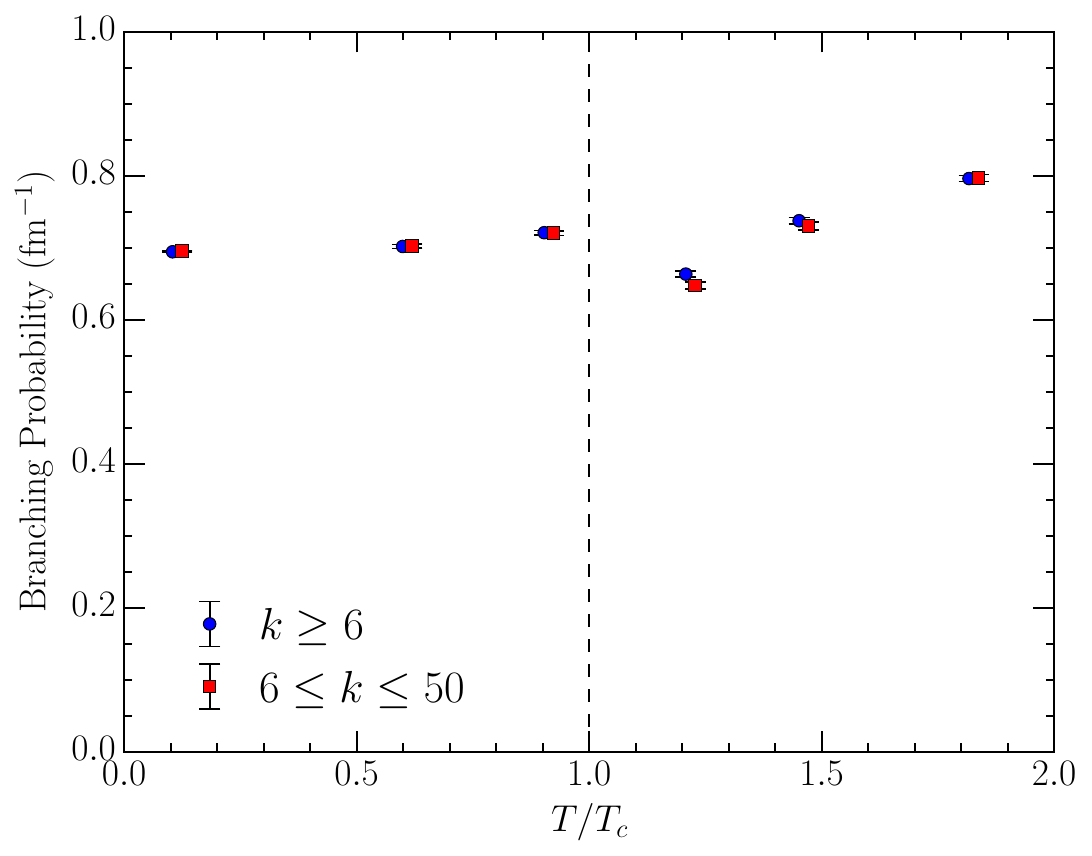}
	\caption{\label{fig:branchingprobability} The long-range ($k\geq 6$) branching probability of a vortex line as a function of temperature, estimated by performing an exponential fit to the distribution of branching point separations. There is a mild drop in the branching probability across the phase transition, which subsequently increases to a value markedly exceeding the below $T_c$ values. Including a cutoff $k\leq 50$ to account for low statistics at large distances is seen to have minimal effect on the results.}
\end{figure}

To account for the $T>T_c$ behaviour discussed above, we perform the fit at all temperatures only for separations $k>5$ and exclusively in temporal slices. This procedure is carried out on 100 bootstrap ensembles, allowing an uncertainty to be placed on the final probability. For each bootstrap sample, Poisson square-root errors are placed on the histogram bin counts for incorporation into the least-squares fit. The fits for each ensemble are shown on a logarithmic scale in Fig.~\ref{fig:bpsepfits},
and the resulting evolution of the branching probability is presented in Fig.~\ref{fig:branchingprobability}.

The fits of Fig.~\ref{fig:bpsepfits} are observed to describe the data well. The suppression of short-distance clustering at high temperatures is abundantly clear. The lower statistics at large distances is especially pronounced for the higher-temperature ensembles, which have less statistics by default. For completeness, we reperform the fits exclusively including up to a maximum separation of $k_\mathrm{max}=50$, which is additionally shown in Fig.~\ref{fig:branchingprobability}. Incorporating this cutoff is seen to have a visible but ultimately insignificant effect, justifying the original fits.

It is expected that the branching probability scales with the lattice spacing \cite{Branching}. This can be intuitively understood by the fact vortex and branching point densities are known to exhibit scale invariance as $a\to 0$. Thus if one were to decrease $a$, the position of vortices would be resolved to greater precision, but the physical distance between branching points would be unchanged. In other words, the probability of branching in a given elementary cube would decrease proportionally to the lattice spacing. For this reason, Fig.~\ref{fig:branchingprobability} shows a physical ``probability per unit length" by dividing the value obtained through the fits by our lattice spacing of $a=0.1\,$fm.

The branching probability displays the same general trends as the previous vortex quantities investigated, including the mild increase approaching $T_c$ from below, the drop through the phase transition and the steady climb in value as the temperature rises above $T_c$. That being said, the cut in probability at the critical temperature is much less pronounced here, and its value is then amplified to notably exceed below $T_c$ levels at our highest temperature.

It is prescient to compare this behaviour to the linear branching point density (Fig.~\ref{fig:linearbranchingdensity}), where if a constant branching probability held over all separations these measures would be expected to coincide. Instead, the probability obtained through the fits is found to consistently sit above the linear density at all temperatures, consistent with the previous finding at zero temperature \cite{StructureDynamical}.

This is a consequence of incorporating the short-distance physics into the naive linear branching density, incompatible with the constant probability picture at large separations. Of utmost interest is the high-temperature behaviour, where although the linear density never exceeds the below $T_c$ values over our temperature range, the branching probability does by a considerable margin. That is to say, the long-distance probability of a vortex undergoing branching is greater at very high temperatures compared to low temperatures. This conclusion was obscured in the linear density, which highlights the importance of separating out the short-distance behaviour in providing a comprehensive assessment of branching point geometry.

\section{Conclusion} \label{sec:conclusion}
In this paper we have explored the evolution of pure-gauge SU(3) centre vortex structures with temperature. Initially, a qualitative discussion was presented by visualising the vortex structure in temporal and spatial slices of the lattice. This established a shift in the centre vortex sheet to principally align with the short temporal dimension in the deconfined phase. As a consequence, the vortex cluster is observed to be primarily frozen between temporal slices of the lattice. By aligning with the temporal axis, the vortex sheet rarely cuts through space-time plaquettes above $T_c$. This is reflected in spatial slices, where the majority of vortices pierce space-space plaquettes and are therefore oriented in the temporal dimension. In other words, this manifests as many short vortex lines chiefly parallel to the temporal axis.

An array of vortex statistics was calculated to investigate additional intrinsic changes to the vortex structure. These statistics include the vortex density, and volume and linear branching point densities. All of these measures exhibit the same overarching trend. Crossing the phase transition, each quantity experiences a cut in value in both temporal and spatial slices, though this was less pronounced in temporal slices. These values in the temporal slices subsequently increase as the temperature continues to rise above $T_c$, returning to below $T_c$ levels by the highest temperature considered $T/T_c\approx 1.827$.

Decomposing the density in spatial slices to consider space-space and space-time plaquettes independently revealed a slight preference of the vortex sheet to pierce space-time plaquettes as the temperature nears $T_c$. We propose this is a necessity as the vortex sheet prepares to align with the temporal dimension. Other peculiar findings, such as an increase in each of the aforementioned statistics whilst approaching the phase transition from below, and a major suppression of secondary clusters above $T_c$, warrant further investigation into their cause. This could be achieved by focusing on a narrower range of temperatures near $T_c$, or utilising finer lattices to study the scaling behaviour of these trends.

Calculating a linear branching density averaged over spatial and temporal slices revealed constant (but distinct) values on both sides of the phase transition. This is in spite of the strong asymmetry between the space and time slices above $T_c$. This embodies a degree of ``locking" in the vortex structure, in which continued changes in spatial slices of the lattice must be counteracted by an opposite trend in temporal slices. This opens the possibility of an underlying symmetry connected with the vortex topology which results in the average linear branching probability becoming an effective constant. In future work, it will be interesting to investigate this quantity over an extended temperature range, both nearer to and farther from $T_c$, to ascertain the extent to which this locking persists in the deconfined phase.

Finally, we explored changes to the inherent distribution of branching points throughout the vortex clusters with temperature. Here, we find a tendency of branching points to cluster at short separations at low temperatures is lost as we move through the phase transition. At large separations, their distribution can nonetheless be described by a constant branching probability. This long-distance probability is decidedly greater at very high temperatures compared to low temperatures. Thus, branching is enhanced at very high temperatures.

By analysing other topological aspects of the gauge fields, it was proposed in Ref.~\cite{AboveTcEvolutionII} that full QCD experiences a second phase transition at a higher temperature $T\approx 2\,T_c$. This was based on the behaviour of the Polyakov loop in different nontrivial representations of SU(3) and the topological index of gauge field configurations. Although our results on centre vortex structure do not evince any evidence for such a second transition, this could be due to the limited temperature range considered or the nature of pure gauge theory compared to full QCD. There is the possibility for further investigation in this regard by utilising the finer resolution provided by an anisotropic lattice to consider a greater range of temperatures.

Future work will extend this analysis to include the effect of dynamical fermions on centre vortices at finite temperature. This has previously been conducted at zero temperature \cite{StructureDynamical}, where many quantities of interest, such as the number of secondary clusters, vortex and branching point densities, and the degree of clustering at short separations are observed to significantly increase. Hence it will be interesting to see the extent to which the findings on vortex geometry through the phase transition presented herein are impacted (or otherwise) by the inclusion of dynamical fermions.

\begin{acknowledgments}
It is a pleasure to thank Prof.\ Chris Allton for many interesting and thought-provoking discussions on quantifying the behaviour of centre vortices at finite temperature, and James Biddle for his contributions in analysing vortex geometry that brought about this research. This work was supported with supercomputing resources provided by the Phoenix High Performance Computing (HPC) service at the University of Adelaide. This research was undertaken with the assistance of resources and services from the National Computational Infrastructure (NCI), which is supported by the Australian Government. This research was supported by the Australian Research Council through Grant No. DP210103706. W.~K.\ was supported by the Pawsey Supercomputing Centre through the Pawsey Centre for Extreme Scale Readiness (PaCER) program.
\end{acknowledgments}

\appendix

\section{Exponential and geometric distributions} \label{app:expgeo}
In this appendix we derive the connection between the exponential and geometric distributions used to extract an estimate of the branching probability from our exponential fits, Eq.~(\ref{eq:exp}). If a continuous random variable $Y$ follows an exponential distribution, $Y\sim\operatorname{Exp}\,(\lambda)$, it has the (normalised) probability density
\begin{equation}
	f(x;\lambda) = \lambda\,e^{-\lambda x} \,.
\end{equation}
The discrete random variable $X=\lfloor Y\rfloor + 1$, supported on the set $\{1,\,2,\,3,\,\hdots\}$, then follows a geometric distribution,
\begin{align}
	\begin{split}
		\Pr\,(X=k) &= \Pr\,(k-1\leq Y < k) \\
		&= \int_{k-1}^k dx\,\, \lambda\,e^{-\lambda x} \\
		&= e^{-\lambda (k-1)} - e^{-\lambda k} \\
		&= \left(1 - e^{-\lambda}\right) e^{-\lambda (k-1)} \\
		&= \left(1 - e^{-\lambda}\right) \left[1 - (1 - e^{-\lambda})\right]^{k-1} \,.
	\end{split}
\end{align}
This coincides with the formula for a geometric distribution, Eq.~(\ref{eq:geodist}), with probability of success
\begin{equation}
	p = 1-e^{-\lambda} \,.
\end{equation}
This verifies Eq.~(\ref{eq:probexp}) for the branching probability, where the random variable $X$ is identified with the branching point separations. Note that in our fits, we necessarily included an unconstrained normalisation parameter $\zeta$ to accommodate the non-exponential behaviour at short separations. However, this has no effect on the above probability, but rather just carries through to an equivalent normalisation factor in the geometric distribution.

\bibliography{main}

\clearpage

\input{supp}

\end{document}

%% file: supp.tex
\title{Supplemental Material: Centre vortex geometry at finite temperature}
\begin{abstract}
	This supplementary document provides animations of the centre vortex structure over temporal and spatial slices of the lattice, expanding on the static images in the main text to build a complete understanding of the four-dimensional structure at finite temperature. The animations are produced using the \texttt{animate} package in \LaTeX. To play the animations, readers must open this document in any of Acrobat Reader, KDE Okular, PDF-XChange or Foxit Reader.
	
	The controls for the animations are located below their thumbnails. From left to right, these are: stop and go to first frame, step backwards one frame, play backwards, play forwards, step forwards one frame, and stop and go to last frame. Simply click on the desired control to interact with the animation. After clicking either play button, they will be replaced with a pause button which can subsequently be used to stop the animation.
\end{abstract}

\makeatletter
\@booleanfalse\preprint@sw
\makeatother
\maketitle

\setcounter{figure}{0}
\renewcommand\thefigure{S-\arabic{figure}}
\renewcommand{\theHfigure}{Supplement.\thefigure}

\begin{figure*}
	\centering
	\animategraphics[loop,nomouse,width=0.59\linewidth,controls={play,stop,step}]{4}{Nt12_TemporalAnimation/t}{1}{12}
	
	\vspace{1.2em}
	
	\animategraphics[loop,nomouse,width=0.58\linewidth,controls={play,stop,step}]{4}{Nt12_SpatialAnimation/x}{1}{32}
	\caption{\label{fig:Nt12Animation} Animations of the centre vortex structure in temporal slices (\textbf{top}) and spatial slices (\textbf{bottom}) of the lattice below the critical temperature at $T/T_c = 0.609$.}
\end{figure*}

\begin{figure*}
	\centering
	\animategraphics[loop,nomouse,width=0.59\linewidth,controls={play,stop,step}]{4}{Nt8_TemporalAnimation/t}{1}{8}
	
	\vspace{1.2em}
	
	\animategraphics[loop,nomouse,width=0.58\linewidth,controls={play,stop,step}]{4}{Nt8_SpatialAnimation/x}{1}{32}
	\caption{\label{fig:Nt8Animation} Animations of the centre vortex structure in temporal slices (\textbf{top}) and spatial slices (\textbf{bottom}) of the lattice below the critical temperature at $T/T_c = 0.913$.}
\end{figure*}

\begin{figure*}
	\centering
	\animategraphics[loop,nomouse,width=0.59\linewidth,controls={play,stop,step}]{4}{Nt6_TemporalAnimation/t}{1}{6}
	
	\vspace{1.2em}
	
	\animategraphics[loop,nomouse,width=0.58\linewidth,controls={play,stop,step}]{4}{Nt6_SpatialAnimation/x}{1}{32}
	\caption{\label{fig:Nt6Animation} Animations of the centre vortex structure in temporal slices (\textbf{top}) and spatial slices (\textbf{bottom}) of the lattice above the critical temperature at $T/T_c = 1.218$.}
\end{figure*}

\begin{figure*}
	\centering
	\animategraphics[loop,nomouse,width=0.59\linewidth,controls={play,stop,step}]{4}{Nt5_TemporalAnimation/t}{1}{5}
	
	\vspace{1.2em}
	
	\animategraphics[loop,nomouse,width=0.58\linewidth,controls={play,stop,step}]{4}{Nt5_SpatialAnimation/x}{1}{32}
	\caption{\label{fig:Nt5Animation} Animations of the centre vortex structure in temporal slices (\textbf{top}) and spatial slices (\textbf{bottom}) of the lattice above the critical temperature at $T/T_c = 1.461$.}
\end{figure*}

\begin{figure*}
	\centering
	\animategraphics[loop,nomouse,width=0.59\linewidth,controls={play,stop,step}]{4}{Nt4_TemporalAnimation/t}{1}{4}
	
	\vspace{1.2em}
	
	\animategraphics[loop,nomouse,width=0.58\linewidth,controls={play,stop,step}]{4}{Nt4_SpatialAnimation/x}{1}{32}
	\caption{\label{fig:Nt4Animation} Animations of the centre vortex structure in temporal slices (\textbf{top}) and spatial slices (\textbf{bottom}) of the lattice above the critical temperature at $T/T_c = 1.827$.}
\end{figure*}